\documentclass[twocolumn,notitlepage,nofootinbib,aps,prl,amsmath,amssymb,superscriptaddress]{revtex4-1}

\usepackage{amsmath,amsfonts,amssymb}
\usepackage[usenames]{color}
\usepackage{hyperref}
\usepackage{graphicx}

\newcommand{\be}{\begin{equation}}
\newcommand{\ee}{\end{equation}}
\newcommand{\bea}{\begin{eqnarray}}
\newcommand{\eea}{\end{eqnarray}}
\newcommand{\rr}{\mathbf{r}}
\newcommand{\kk}{\mathbf{k}}
\newcommand{\qq}{\mathbf{q}}

\newcommand{\dd}{\mathrm{d}}

\newcommand{\bb}[1]{\left( #1 \right)}

\usepackage[normalem]{ulem}

\usepackage{xcolor}

\begin{document}
\title{{Higgs} oscillations in a unitary Fermi superfluid} 

\author{P. Dyke}
\affiliation{Optical Sciences Centre, ARC Centre of Excellence in Future Low-Energy Electronics Technologies, Swinburne University of Technology, Melbourne 3122, Australia}
\author{S. Musolino}
\affiliation{Universit\'e C\^ote d'Azur, CNRS, Institut de Physique de Nice, 06200 Nice, France}
\author{H. Kurkjian}
\affiliation{Laboratoire de Physique Th\'eorique, Universit\'e de Toulouse, CNRS, UPS, France}
\author{D. J. M. Ahmed-Braun}
\affiliation{Eindhoven University of Technology, P.O. Box 513, 5600 MB Eindhoven, The Netherlands}
\author{A. Pennings}
\affiliation{Optical Sciences Centre, ARC Centre of Excellence in Future Low-Energy Electronics Technologies, Swinburne University of Technology, Melbourne 3122, Australia}
\author{I. Herrera}
\affiliation{Optical Sciences Centre, ARC Centre of Excellence in Future Low-Energy Electronics Technologies, Swinburne University of Technology, Melbourne 3122, Australia}
\author{S. Hoinka}
\affiliation{Optical Sciences Centre, ARC Centre of Excellence in Future Low-Energy Electronics Technologies, Swinburne University of Technology, Melbourne 3122, Australia}
\author{S. J. J. M. F. Kokkelmans}
\affiliation{Eindhoven University of Technology, P.O. Box 513, 5600 MB Eindhoven, The Netherlands}
\author{V. E. Colussi}
\affiliation{Pitaevskii BEC Center, CNR-INO and Dipartimento di Fisica, Università di Trento, 38123 Trento, Italy}
\affiliation{Infleqtion, Inc., 3030 Sterling Circle, Boulder, CO 80301, USA}
\author{C. J. Vale}
\affiliation{Optical Sciences Centre, ARC Centre of Excellence in Future Low-Energy Electronics Technologies, Swinburne University of Technology, Melbourne 3122, Australia}
\affiliation{CSIRO, Research Way, Clayton 3168, Australia}

\begin{abstract}  
{Symmetry-breaking phase transitions are central to our understanding of states of matter. When a continuous symmetry is spontaneously broken, new excitations appear that are tied to fluctuations of the order parameter. In superconductors and fermionic superfluids, the phase and amplitude can fluctuate independently, giving rise to two distinct collective branches. However, amplitude fluctuations are difficult to both generate and measure, as they do not couple directly to the density of fermions and have only been observed indirectly to date. Here, we excite amplitude oscillations in an atomic Fermi gas with resonant interactions by an interaction quench. Exploiting the sensitivity of Bragg spectroscopy to the amplitude of the order parameter, we measure the time-resolved response of the atom cloud, directly revealing amplitude oscillations at twice the frequency of the gap. The magnitude of the oscillatory response shows a strong temperature dependence, and the oscillations appear to decay faster than predicted by time-dependent BCS theory applied to our experimental setup.}

\end{abstract}




\maketitle

The ability of interacting particles to act collectively underpins many of the remarkable properties of quantum matter. From superfluidity and superconductivity to magnetism and elementary particles, order parameters and their fluctuations govern a wide variety of collective quantum phenomena~\cite{Sachdev11}. Phase transitions characterized by a complex bosonic order parameter are generally accompanied by the emergence of two distinct collective excitations. Phase fluctuations that manifest as sound waves in neutral systems \cite{Bogoliubov47} become massive in the presence of long-range interactions \cite{Anderson58,Repplinger2023}, while  amplitude (or strictly-speaking modulus) fluctuations are always gapped. This behavior is reminiscent of the Higgs field \cite{Higgs64} in high-energy physics, whose phase is responsible for mass acquisition via the Anderson-Higgs mechanism, and whose amplitude remains electrically neutral and becomes the Higgs boson. The analogy relies on the iconic ``Mexican hat'' potential \cite{Pekker15,Tsuji2020}, governing the dynamics of these complex bosonic fields. Among the systems where an effective action with this form emerges in non-relativistic matter are Bose gases near the superfluid-Mott insulator transition~\cite{Bissbort11,Endres12}, spinor Bose-Einstein condensates (BECs)~\cite{Hoang16}, atoms in optical cavities~\cite{Leonard17}, dipolar gases in the supersolid phase~\cite{Hertkorn19} and antiferromagnetic materials~\cite{Jain17}.

The analogy with the Higgs field is often extended to the order parameter $\Delta$ of fermionic pair condensates \cite{Matsunaga2013,Volovik2016,Behrle18}. This case however, is more subtle as the dynamics of $|\Delta|$ result from (pair-breaking) bi-excitations of the fermionic quasiparticles. Unlike phase fluctuations, which obey superfluid hydrodynamics \cite{Khalatnikov}, amplitude fluctuations \textit{cannot} be modelled by a low-energy effective action such as the Mexican hat potential \cite{Cea15}, and remain an intrinsically many-body phenomenon, with unique phenomenology.

The microscopic description of a Bardeen-Cooper-Schrieffer (BCS) superconductor/superfluid shows that there exists a collective amplitude mode within the pair-breaking continuum \cite{Popov1976},  which persists even in presence of amplitude-phase coupling~\cite{Kurkjian19}. In the zero-momentum limit, the spectral weight of the amplitude mode vanishes, yet \textit{amplitude oscillations} still occur due to the presence of a non-analytic singularity in the amplitude response function. 
Within a mean-field approximation, the frequency of these amplitude oscillations is set at twice the gap in the fermionic excitation spectrum \cite{Scott12}, and the oscillations decay according to a power-law with an exponent that changes at the transition from BCS to the BEC regime~\cite{Volkov74,Yuzbashyan06,Gurarie09,Liu16,Tokimoto17}. In the regime of nonlinear excitations, other asymptotic behaviours become possible, including persistent oscillations~\cite{Yuzbashyan15,LewisSwan21,Young23}. 

Nonlinear amplitude oscillations have been recorded through third harmonic generation in  BCS~\cite{Matsunaga2013} and cuprate~\cite{Katsumi18,Chu20} superconductors, and a dressed amplitude mode has been observed in charge density wave~\cite{Klein1980,Sacuto2014,Measson2019} superconductors. The case of neutral Fermi gases is \textit{a priori} favorable since the strength $|\Delta|$ of the pair condensate can be accessed directly, either by tuning the interaction strength \cite{Jin2005}, or via radio-frequency (rf)  coupling to a third internal state \cite{Ketterle2008}. To date, a broad spectroscopic peak was reported around the threshold of the pair-breaking continuum \cite{Jin2005,Behrle18}, but the spectral resolution was too low to unambiguously identify the singularity responsible for amplitude oscillations. Modulated interactions have previously been used to study the dynamics of pair condensation~\cite{Zwierlein05,Harrison21,Dyke21}.
%


Here, we directly observe amplitude oscillations in an ultracold atomic Fermi condensate with 
resonant interactions. We excite the oscillations by a uniform (zero-momentum) quench of the 
interactions using a magnetic Feshbach resonance. We probe the ensuing out-of-equilibrium 
dynamics {using high-momentum Bragg scattering, tuned to resonantly excite condensed pairs, 
which is highly sensitive to variations of the order parameter. Our real-time experiment allows 
us to characterize the frequency, magnitude and decay of the oscillations. Comparing to predictions from time-dependent BCS theory, our experiment confirms oscillations occur at twice the ($2\Delta$) and show qualitative agreement on the temperature dependence of the oscillation magnitude, with a reduction as the number of condensed pairs decreases near the critical temperature $T_c$ \cite{Volkov74}. The observed oscillations at unitarity decay faster than predicted by BCS theory, even when experimental effects such as inhomogeneous broadening are taken into account.}

\begin{figure}
\includegraphics[width=\columnwidth]{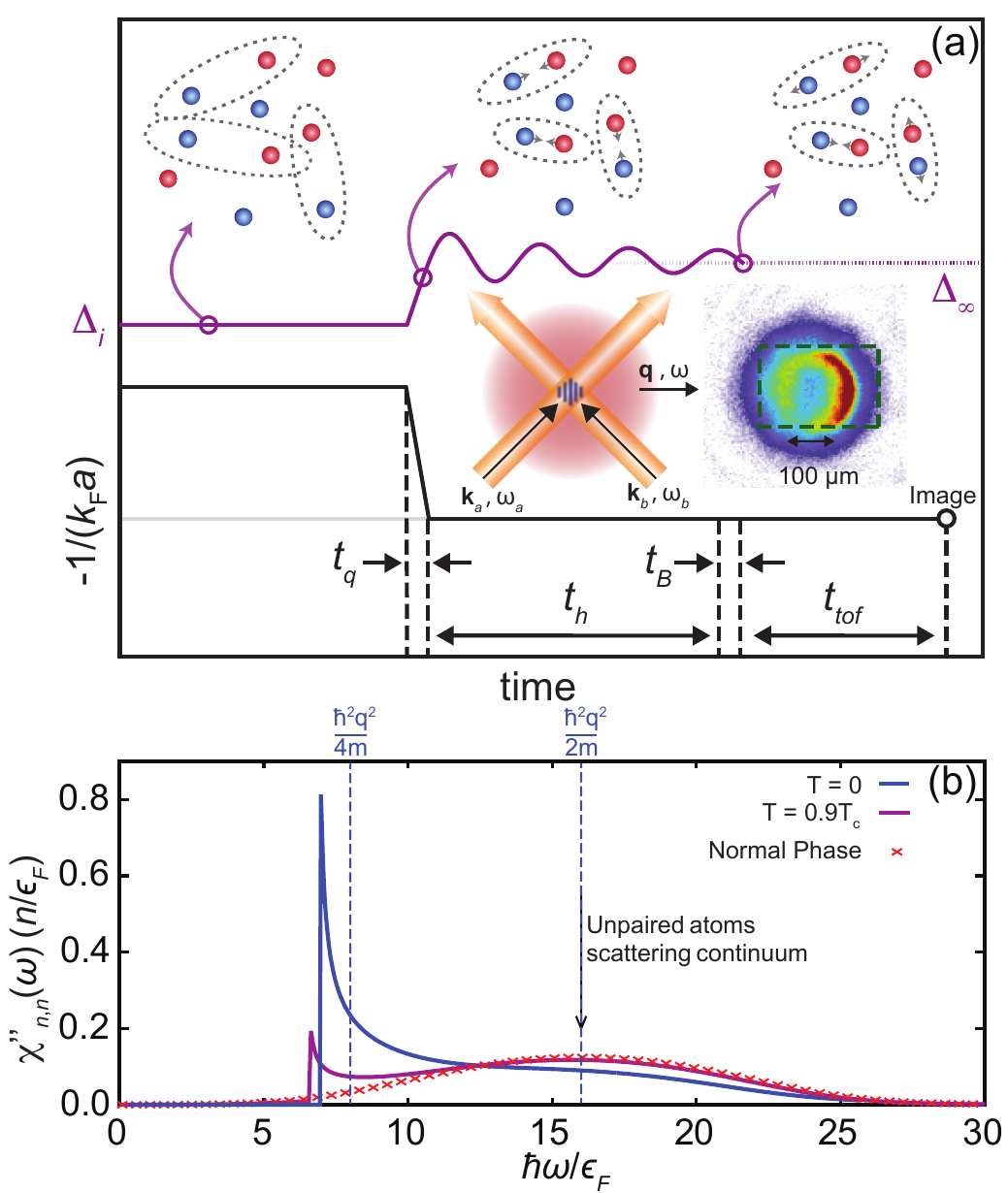}
\caption{Excitation and detection of amplitude oscillations in a paired Fermi superfluid. (Top panel) Pairs of fermions (dashed ellipses) initially at equilibrium are excited by a rapid variation of the interatomic interactions in a time $t_q = 50 \, \mu$s. After a variable hold time $t_h$, we measure the Bragg response of the nearly-uniform central region. The momentum imparted by the pulse is accessed through the center-of-mass displacement in time-of-flight images~\cite{suppmat}. The quench projects the pairs into a superposition of the more tightly bound ground state and the continuum of fermionic biexcitations, with energies $2\epsilon_k$. The pairing field thus begins oscillating, triggering oscillations of the order parameter (purple curve). The continuum edge at $2\Delta \equiv 2 \min_k(\epsilon_k)$ sets the frequency of the oscillations, which attenuate over time due to the spread of the excited state wave function over energies $2\epsilon_k$, eventually stabilizing at $\Delta_{\infty}$. At nonzero temperatures, the superfluid pairs are surrounded by a thermal cloud of unpaired atoms (isolated blue and red dots), reducing the spectral weight of the amplitude oscillations. (Bottom panel) 
The imaginary part of the density-density response function in the Random-Phase Approximation (RPA) and for the large pair center-of-mass momentum $q$ 
used in our Bragg spectroscopy. The energy and magnitude of the peak
at the dissociation threshold $\hbar\omega_{\rm th}=\sqrt{4\Delta^2+(\hbar^2q^2/4m-\mu)^2}$ varies with $\Delta$ during the post-quench evolution, which makes our Bragg measurement sensitive to the amplitude oscillations. \label{fig1v2}}
\end{figure}

Our experimental protocol is depicted in Fig.~\ref{fig1v2}~\cite{Scott12}. An ultracold gas of fermionic $^6$Li atoms is prepared in a balanced mixture of two spin states, initially at thermal equilibrium. Elastic collisions between atoms in these states can be tuned by an external magnetic field through a broad Feshbach resonance~\cite{suppmat}. Interactions are characterised by the dimensionless parameter $1/(k_{\rm F} a)$ where $k_{\rm F} = (3 \pi^2 n)^{1/3}$ is the Fermi wave vector, $n$ is the atomic density and $a$ is the $s$-wave scattering length. The cloud is initially prepared below $T_c$, slightly to the BCS side of the Feshbach resonance ($1/(k_{\rm F} a_i) \approx -0.18 \pm 0.02$). The magnetic field is then ramped to unitarity (where $a \rightarrow \infty$) in a time $t_q = 50 \, \mu$s, too fast for the system to follow adiabatically, creating a superposition of the more strongly paired ground state and the continuum of excited states. As this superposition evolves, the pairing field oscillates at a frequency set by the energy difference between the ground and excited states, leading to Higgs oscillations of the order parameter. 

{According to Refs.}\cite{Volkov74,Gurarie09,Yuzbashyan15} a power-law damping of the oscillations occurs, due to the spread in energy of the lowest lying excited states.
In the BCS (weak-coupling) limit, the lowest energy excitations occur at the Fermi surface, $p \approx \hbar k_{\rm F}$, where the 3D density of excited states is large, and {this small spread in energy leads to oscillations decaying slowly, as $t^{-1/2}$}~\cite{Volkov74}.
In the opposite limit of tightly bound molecules, the dispersion minimum occurs at $p = 0$, where the density-of-state vanishes, as for free particles. The evolution of the excited wave function is thus similar to a 3D ballistic expansion and the overlap with the molecular {ground state} decays as $t^{-3/2}$~\cite{Gurarie09}.

We model these dynamics using time-dependent BCS theory~\cite{Ripka1985}. The initial state of the gas is treated in first approximation as a homogeneous BCS state at nonzero temperature, containing both superfluid pairs and unpaired thermal atoms with a Fermi-Dirac distribution $n_{\rm F}(\epsilon_{\kk,i})=1/(1+\text{exp}(\epsilon_{\kk,i}/k_BT))$, where $\epsilon_{\kk,i}=\sqrt{(\hbar^2k^2/2m-\mu_i)^2+\Delta_i^2}$ is the initial spectrum, $\Delta_i$ and $\mu_i$ the initial gap and chemical potential, respectively. Following the quench, the initial momentum distribution of the atoms $n_{\kk}(t=0)=n_{\kk,i}$ and pair correlation function $c_{\kk}(t=0)=c_{\kk,i}$
are out-of-equilibrium and evolve according to the time-dependent BCS equations:
\begin{eqnarray}
i\hbar \partial_t{n}_\kk &=& \Delta c_\kk^\ast - c_\kk \Delta^\ast,\label{eq:nk_dot}\\
 i\hbar \partial_t{c}_\kk&=& (\hbar^2k^2/m) c_\kk +  (1-2 n_\kk) \Delta \label{eq:ck_dot}
\end{eqnarray}
where a non-linearity is caused by the gap equation $\Delta(t)=g_0\int\dd^3 k c_\kk/(2\pi)^3$ with $g_0$ the coupling constant of the short-range interactions.

For temperatures well below $T_c$, our quench is shallow ($|\Delta_i-\Delta(t)|\ll \Delta_i$), and the cloud remains close to equilibrium. In this limit, the dynamical system \eqref{eq:nk_dot}--\eqref{eq:ck_dot} can be treated within linear response and the time-evolution of $\Delta$ expressed as a Fourier transform of the amplitude-amplitude response function $\chi_{|\Delta||\Delta|}$~\cite{suppmat}:
\begin{equation}
\Delta(t)-\Delta_{\infty}\propto \int_{2\Delta/\hbar}^{+\infty}\frac{\cos \omega t}{ \omega}\chi''_{|\Delta||\Delta|}(\omega)\dd\omega, 
\end{equation}
where the asymptotic value $\Delta_{\infty}=\Delta(t\to+\infty)$ is not necessarily the equilibrium state in this integrable theory.
This frequency integral covers the superposition of all excited states with energy $2\epsilon_k$, giving rise to the collective response of $\Delta(t)$. The gapped BCS spectrum sets the lower bound $2\Delta/\hbar$, and the behavior near this pair-breaking threshold governs the long-time behavior of $\Delta(t)$. In the BCS regime ($\mu_i>0$, which includes unitarity), the amplitude response has a square-root singularity at the continuum edge, $\chi''_{|\Delta||\Delta|}\underset{\omega\to 2\Delta/\hbar}{\propto} 1/\sqrt{\omega-2\Delta/\hbar}$ leading to power-law damped oscillations of the form
\be
\frac{\Delta(t)-\Delta_\infty}{\Delta_i-\Delta_{\infty}}  \underset{t\gg \tau_{\rm F}}=
A_{\rm th}\frac{\cos\bb{2\Delta t/\hbar+{\pi}/{4}}}{\bb{2\Delta t/\hbar}^{\gamma_{\rm th}} }.
\label{Deltat}
\ee
We find that the amplitude $A_{\rm th}$ decreases with temperature, whereas the damping exponent $\gamma_{\rm th}=1/2$ stays constant.
For larger quenches triggering nonlinear dynamics, the oscillatory form \eqref{Deltat} can remain valid but the oscillation frequency $\omega_{\rm H}$ deviates from $2\Delta/\hbar$ \cite{Yuzbashyan15,suppmat}.

{We probe these dynamics using Bragg spectroscopy.} Our experiments use atom clouds confined in an oblate harmonic potential, formed by a combination of optical and magnetic fields~\cite{suppmat}, leading to a non-uniform density distribution. As a consequence the pairing gap $\Delta(\rr)$, set by the local Fermi energy, $E_{\rm F}(\rr) = \hbar^2 (3 \pi^2 n(\rr)^{2/3})/(2m)$, varies with position $\rr$ across the cloud. 
To overcome this, we probe only a small, near-homogeneous volume of the cloud using two-photon Bragg scattering. At the end of the hold time $t_h$, we send in two tightly-focused Bragg lasers (Fig.~\ref{fig1v2}), that intersect in the centre of the trapped cloud, where the density distribution is most uniform~\cite{Hoinka17,Carcy19}. We define the average density in the Bragg volume $\bar{n} = \int \Omega_{\mathrm{Br}}(\rr) n(\rr) \mathrm{d}^3 \rr / \int \Omega_{\mathrm{Br}}(\rr) \mathrm{d}^3 \rr$, where $\Omega_{\mathrm{Br}}(\rr)$ is the spatially dependent two-photon Rabi frequency. In the experiments presented here, we find $\bar{n} =(0.955 \pm 0.018) n_0$, where $n_0$ is the peak density in the trap centre, to be independent of temperature within our experimental resolution~\cite{suppmat}. 
{The remaining small inhomogeneities can be accounted for in our theoretical description within the local density approximation \cite{suppmat}.
They cause an additional damping of the oscillations, as regions oscillating at different frequencies gradually dephase. }

To resonantly excite pairs with zero center-of-mass momentum, we set the frequency difference between the two lasers to half of the atomic recoil ($\hbar\omega_r/2=\hbar^2 q^2/(4m)$)~\cite{Lingham14,Carcy19}. Bragg scattered pairs begin moving with a velocity $\hbar \qq /(2m)$ where $\qq = \kk_a - \kk_b$ is the difference of the wave-vectors of the two Bragg lasers. We use $q \simeq 4 k_{\rm F}$ to ensure that $\hbar\omega$ is large compared to $E_{\rm F}$, and the Bragg pulse duration ($t_B = 50 \, \mu$s) provides good spectral resolution, while remaining 3 to 4 times shorter than the typical oscillation period ($\tau_{\rm H} = 2 \pi / \omega_{\rm H}$) so the oscillations remain visible.  {We estimate that the observed oscillation magnitude is reduced by less than $15\%$ due to this time-averaging \cite{suppmat}.}

The resulting center of mass displacement ${S=}\Delta X_{\mathrm{CoM}}$ following time-of-flight expansion, is proportional to the {momentum transferred} to the atoms by the Bragg lasers~\cite{suppmat}, {hence} to the imaginary part of density-density response function $\chi''_{nn}(\omega_r/2, q=4k_{\rm F}$) \cite{Hoinka17}.
At large $q$, $\chi''_{nn}$ has a sharp peak at the continuum threshold (bottom panel of Fig.~\ref{fig1v2}) which coincides approximately with the pair recoil frequency~\cite{Combescot06,Klimin2020}. Both the height and energy of this peak are sensitive to variations in $\Delta$. When $t_B \ll \tau_{\rm H}$, the Higgs oscillations are approximately stationary during the Bragg pulse and the time-dependent Bragg response can be written as
\be
\chi''_{nn}(\omega,q,t)\approx \chi''_{nn}(\omega,q;\Delta_i)+\frac{\dd\chi''_{nn}}{\dd\Delta}(\Delta(t)-\Delta_i),\label{sensitivity}
\ee
Our Bragg frequency $\omega = \omega_r/2$ sits just on the high energy slope of the threshold peak~\cite{suppmat}, where $\chi''_{nn}$ is very sensitive to variations of $\Delta$. Experimentally, we observe that the Bragg response at $\omega = \omega_r/2$ shows a strong dependence on the condensate fraction, reflecting the temperature dependence of the spectral weight of this threshold peak~\cite{suppmat}.

\begin{figure*}[ht]
    \includegraphics[clip, width = \textwidth]{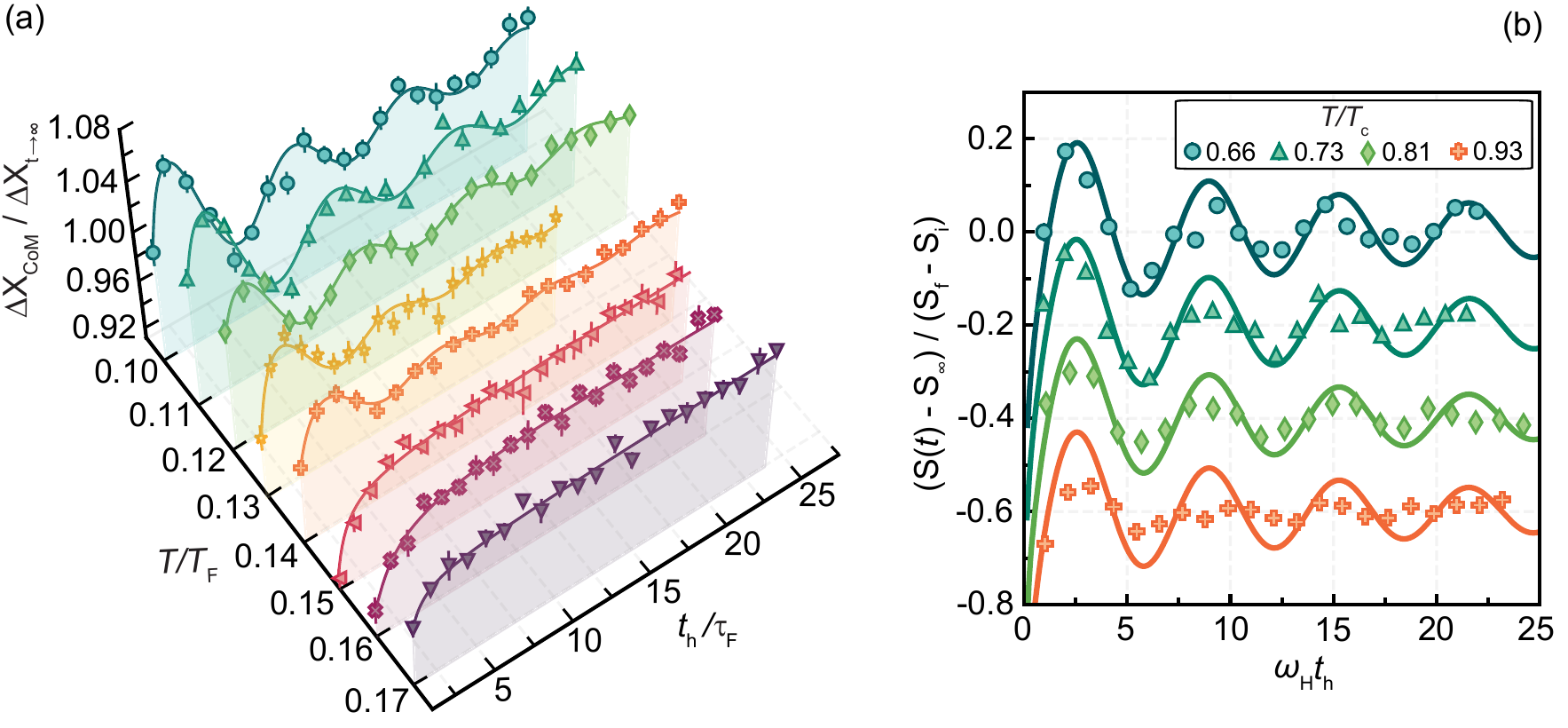}
    \caption{(Color online) (a) Bragg response (centre of mass displacement $S$), relative to the asymptotic response $S_{\infty}$ ($t \rightarrow \infty$), as function of hold time after the quench for a selection of (final) equilibrium cloud temperatures. Points are the experimental measurements and solid lines are fits to the data of a power-law damped sinusoidal function (see text). (b) Comparison with time-dependent BCS theory including experimental effects \cite{suppmat}. The experimental points are shown as a function of $ \omega_{\rm H}t_h$ and $T/T_c$ using the fitted value of $\hbar \omega_{\rm H}/\epsilon_{\rm F}$ and the estimated value $T_{c,i}/T_{\rm F}\simeq0.15$~\cite{Haussmann07} at $1/k_{\rm F}a=-0.18$. The Bragg signal $S(t)-S_\infty$ is scaled to its variation $S_f-S_i$ under an adiabatic sweep of the scattering length, which we measured independently, and the theoretical curves are offset by the delay accumulated during the ramp \cite{suppmat}. The curves for different values of $T$ are vertically offset by $0.2$ for readability.\label{fig2}}
\end{figure*}

Armed with this capability, we use local Bragg scattering as a sensitive, temporally resolved probe fore oscillations of the order parameter. Fig.~\ref{fig2} shows examples of the measured Bragg response, as a function of hold time $t_h$, in units of the local Fermi time $\tau_{\rm F} = \hbar/E_{\rm F}$, for a range of temperatures\footnote{Note that the temperature of the cloud was measured after the quench at unitarity. This will therefore include some heating that occurs due to the non-adiabatic experimental quench, which is not accounted for by BCS theory.}~\cite{suppmat}. A damped oscillation is clear in the Bragg response of the colder clouds, giving a direct signature of the Higgs oscillations. The magnitude of the oscillations decreases for warmer clouds, until non-oscillatory behavior is observed for $T\gtrsim0.15T_{\rm F}$. Also shown are fits of the data to a function of the form ${S}(t) = A_{\rm ex} \, \cos{(\omega_{\rm H} t + \phi)} / t^{\gamma} + S_{\infty}$ where $A_{\rm ex}$, $\omega_{\rm H}$, $\phi$, $\gamma$ and $S_{\infty}$ are fit parameters that characterize the oscillations. 

\begin{figure}[ht]
    \includegraphics[clip, width = 1 \columnwidth]{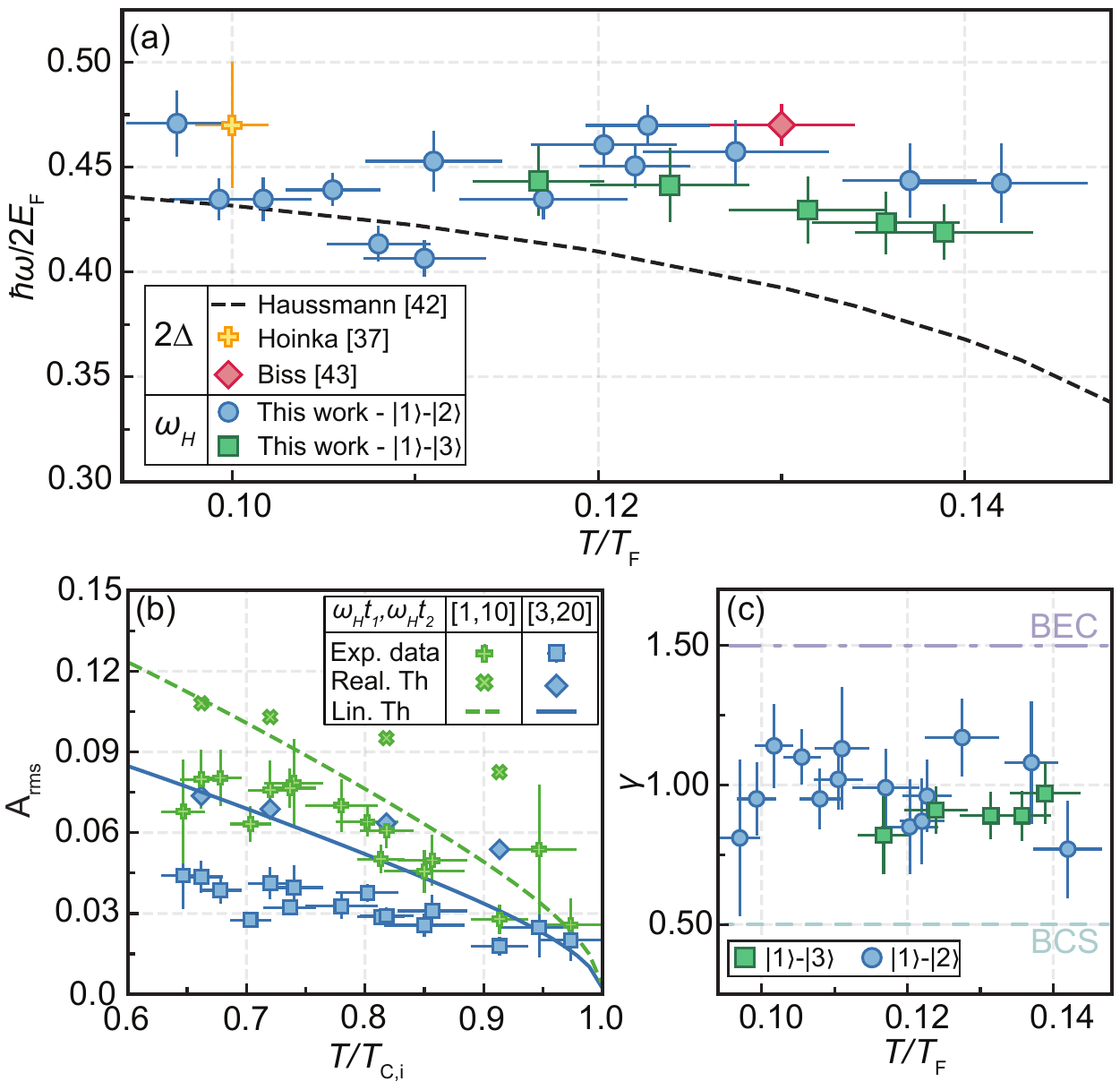}
    \caption{(Color online) (a) Frequency of the Higgs oscillation $\omega_{\rm H}$ versus the normalised temperature $T/T_{\rm F}$ along with previous measurements and a theoretical calculation (dashed line) of $2\Delta$. Blue circles and green squares represent measurements using different combinations of internal states (but for the same interaction quench)~\cite{suppmat}.  (b) The root-mean-square magnitude of the oscillations, measured experimentally (symbols with error bars), predicted analytically from the amplitude response function (solid line), and obtained from a numerical model taking into account experimental imperfections (symbols without error bars) \cite{suppmat}.  (c) The {fitted} damping exponent $\gamma$ of the Higgs oscillation.}
    \label{fig3}
\end{figure}

To compare our experimental measurements to theory, we obtain the asymptotic Bragg response $S_\infty$ ($t\to\infty$), and the separately measure the responses $S_i$ and $S_f$ at thermal equilibrium with the initial and final scattering lengths. From these we construct the ratio $(S(t)-S_\infty)/(S_f-S_i)$, which we directly compare to the theoretical equivalent $(\Delta(t)-\Delta_\infty)/(\Delta_f-\Delta_i)$. The advantage of comparing these quantities is that they do not depend on the experimental sensitivity $\dd\chi_{nn}''/\dd\Delta$ or the offset in the experimental data due to the normal phase response $\chi_{nn}(T>T_c)$, which is not captured in BCS theory. Note the experimental and theoretical temperatures are scaled by the respective critical temperatures of the initial clouds $T_{c,i}$. In Fig.~\ref{fig2}(b) we see good agreement in the dynamics at short times and lower temperatures, however at later times, the experimental signal decays faster than theoretically predicted. 
This is emphasized in Fig.~\ref{fig3}(b) which shows the root-mean-square amplitude $A_{\rm rms}=\sqrt{\frac{1}{t_2-t_1}\int_{t_1}^{t_2}\dd t(S(t)-S_\infty)^2/(S_f-S_i)^2}$, a quantity which does not depend on any of the fitted parameters apart from $S_\infty$. 
While the theory overestimates the magnitude of the oscillations by only 10-20\% in the short time window $1\leq t/\tau_{ H}\leq10$, the overestimate grows to roughly a factor of two at later times $3\leq t/\tau_{ H}\leq20$. Although the quicker decay of the experimental signal may be due to experimental effects other than those we have taken into account in our realistic theory \cite{suppmat},
we note that the prediction of a slow, power-law decay is based on integrable, collisionless theories~\cite{Yuzbashyan15} and may be violated at long times, in particular at times comparable to the quasiparticle collision time \cite{Volkov74}.

From the fits to the experimental data we extract the oscillation frequency $\omega_{\rm H}$ and damping exponent $\gamma$. Fig.~\ref{fig3}(a) shows $\hbar \omega_{\rm H}/2E_{\rm F}$ versus temperature for data points taken in the $|F=1/2, m_F = \pm 1/2\rangle$ hyperfine states (= $|1\rangle$ - $|2\rangle$, blue circles) and $|F=1/2, m_F = +1/2\rangle$ - $|F=3/2, m_F = -3/2\rangle$  hyperfine states (= $|1\rangle$  - $|3\rangle$, green squares) and confronts the data to a selection of previous measurements and calculations of the pairing gap $\Delta$. Theoretically, we expect $\hbar \omega_{\rm H}$ to provide a lower bound on $2\Delta$, and to approach this value at low temperatures when our quench is in the shallow regime. Our measurements lie mostly in the range $0.4 \lesssim \hbar\omega_{\rm H}/2E_{\rm F} \lesssim 0.5$. At low temperature,  they are in good agreement with previous measurements of $2\Delta$~\cite{Ketterle2008,Hoinka17,Moritz2022}, as well as beyond mean-field predictions~\cite{art:Haussmann_2009,Pisani18} and quantum Monte-Carlo calculations~\cite{Magierski2006,Carlson2008}. {Although $\Delta$ is expected to vanish with a critical exponent of $\nu\simeq0.62$ at $T_c$ \cite{Wetterich2010}, we do not observe a noticeable reduction of $\omega_{\rm H}$ in the temperature range we probe.}

Fig.~\ref{fig3}(c) shows the fitted damping exponents $\gamma$ which all lie close to unity. While the uncertainties in $\gamma$ are relatively large, our measurements are not consistent with either the BEC or BCS exponents and display no obvious temperature dependence. The average of our measured damping coefficients is $\bar{\gamma} = 0.98\pm0.15$. {This is significantly above the theoretical prediction of $\gamma_{\rm th} = 0.50 \pm 0.02$~\cite{suppmat} where we take  into account the inhomogeneous density and the finite experimental time window. These effects lead to compensating shifts  on the BCS prediction $\gamma_{\rm th}=1/2$, resulting in a correction that is small compared to the difference between BCS and BEC limits.}

We note that fitting an exponentially decaying cosine function to the experimental data gives a statistically indistinguishable quality of fit such that we cannot rule out exponential decay or that $\gamma$ is affected by other ergodic processes such as quasiparticle collisions. {In the vicinity of $T_c$, the local density approximation may also break down for describing delocalised pairs. Effects of the inhomogeneity of the cloud may thus become enhanced even in the nearly-uniform region probed by our Bragg beams.}

Fifty years after their prediction~\cite{Volkov74}, we present the direct observation of amplitude oscillations in a weakly-excited Fermi superfluid. Using Bragg spectroscopy we probe the real-time dynamics in a unitary Fermi gas, in qualitative agreement with time-dependent BCS theory at low temperatures. Our work opens a wide avenue of research, with possible direct extensions to the BCS and BEC regimes, different quench regimes~\cite{Yuzbashyan15} or dynamical crossings of the phase transition~\cite{Zwierlein05,Harrison21,Dyke21}. Our work also opens pathways to investigate ergodic evolution and the possibility of achieving pre-equilibrated states in strongly interacting quantum matter.

\begin{acknowledgments}
We thank Y. Castin, F. Dalfovo, M. Davis, N. Navon, C. Sa de Melo, S. Stringari, and M. Zwierlein for valuable discussions and comments on the manuscript. This work was supported by the ARC Centre of Excellence for Future Low-Energy ELectronics Technologies.  V. Colussi acknowledges financial support from Provincia Autonoma di Trento, the Italian MIUR under the PRIN2017 projectCEnTraL and the National Science Foundation under Grant No. NSF PHY-1748958.  S. Musolino acknowledges funding from the ANR-21-CE47-0009 Quantum-SOPHA project. D.J.M. Ahmed-Braun and S.J.J.M.F. Kokkelmans acknowledge financial support from the Dutch Ministry of Economic Affairs and Climate Policy (EZK), as part of the Quantum
Delta NL programme, and by the Netherlands Organisation for Scientific Research (NWO) under Grant No.680.92.18.05 and No.680.47.623.
\end{acknowledgments}

\bibliographystyle{apsrev4-1} 
\bibliography{main_biblio}

\end{document}


\title{Supplementary Material}

\author{P. Dyke}
\affiliation{Optical Sciences Centre, ARC Centre of Excellence in Future Low-Energy Electronics Technologies, Swinburne University of Technology, Melbourne 3122, Australia}
\author{S. Musolino}
\affiliation{Universit\'e C\^ote d'Azur, CNRS, Institut de Physique de Nice, 06200 Nice, France}
\author{H. Kurkjian}
\affiliation{Laboratoire de Physique Th\'eorique, Universit\'e de Toulouse, CNRS, UPS, France}
\author{D. J. M. Ahmed-Braun}
\affiliation{Eindhoven University of Technology, P.O. Box 513, 5600 MB Eindhoven, The Netherlands}
\author{A. Pennings}
\affiliation{Optical Sciences Centre, ARC Centre of Excellence in Future Low-Energy Electronics Technologies, Swinburne University of Technology, Melbourne 3122, Australia}
\author{I. Herrera}
\affiliation{Optical Sciences Centre, ARC Centre of Excellence in Future Low-Energy Electronics Technologies, Swinburne University of Technology, Melbourne 3122, Australia}
\author{S. Hoinka}
\affiliation{Optical Sciences Centre, ARC Centre of Excellence in Future Low-Energy Electronics Technologies, Swinburne University of Technology, Melbourne 3122, Australia}
\author{S. J. J. M. F. Kokkelmans}
\affiliation{Eindhoven University of Technology, P.O. Box 513, 5600 MB Eindhoven, The Netherlands}
\author{V. E. Colussi}
\affiliation{Pitaevskii BEC Center, CNR-INO and Dipartimento di Fisica, Università di Trento, 38123 Trento, Italy}
\affiliation{Infleqtion, Inc., 3030 Sterling Circle, Boulder, CO 80301, USA}
\author{C. J. Vale}
\affiliation{Optical Sciences Centre, ARC Centre of Excellence in Future Low-Energy Electronics Technologies, Swinburne University of Technology, Melbourne 3122, Australia}
\affiliation{CSIRO, Research Way, Clayton 3168, Australia}
\maketitle

\section{Experimental protocol}
\subsection{Sample Preparation}\label{sec:preparation}

In our experiments, we cool a balanced mixture of fermionic $^6$Li atoms in two hyperfine states inside a 100 W, 1075 nm single beam optical dipole trap. 
Initially, we performed the quench experiments using an equal mixture of the $| F=1/2, m_{{F}} = + 1/2 \rangle -| F=3/2, m_{{F}} = -3/2 \rangle$ ($\equiv | 1 \rangle - | 3 \rangle$, green squares in Fig.~3, main text). The reduced width of this Feshbach resonance facilitated a slight increase in the rate of change of the interaction strength from $1/k_F a_i = -0.18$ to unitarity, as the necessary change in the magnetic field is reduced. However, the temperature range was severely restricted due to heating during the preparation of the $| 1 \rangle - | 3 \rangle$ mixture. To achieve lower temperatures and, consequently, a broader temperature range, we later employed the $| F=1/2, m_{{F}} = + 1/2 \rangle -| F=1/2, m_{{F}} = -1/2 \rangle$ ($\equiv | 1 \rangle - | 2 \rangle$ blue circles in Fig.~3, main text) Feshbach resonance.

Degeneracy is reached through evaporative cooling by smoothly lowering the trap laser power at the magnetic field where the $\it{s}$-wave scattering length diverges, $\it{a}$ $\rightarrow$ $\infty$. Subsequently, the atom cloud is transferred to an oblate harmonic potential, formed by a combination of optical and magnetic fields. The oblate trap is formed between two anti-nodes of a cylindrically focused, 532-nm (blue-detuned), TEM$_{01}$ mode laser beam~\cite{Smith05, Rath10, Dyke16}, where the two anti-nodes are separated by $\approx$ 90~$\mu$m propagating along the y-direction and the 1/e$^2$ radius in the x-direction is $\approx$ 1.0~mm. The optical potential produces the confinement in the $z$ direction and a very weak anti-confinement in the $x$-$y$ plane. The residual magnetic field curvature from the Feshbach coils provides highly harmonic and cylindrically symmetric confinement in the $x$-$y$ plane, which dominates the anti-trapping of the optical potential. The measured trapping frequencies are $\omega_z/2\pi$ = 105 Hz(1) and ($\omega_x, \omega_y$) = 2$\pi$ $\times$ (24.5, 24.5) Hz at a magnetic field of $B$ = 832.2 G (corresponding to the $| 1 \rangle - | 2 \rangle$ Feshbach resonance~\cite{Zurn13}) and $\omega_z$/2$\pi$ = 103 Hz(1) and $\omega_x$,$\omega_y$ = 2$\pi$ $\times$ (22,22) Hz at a magnetic field of B = 689.89 G (corresponding to the $|1\rangle$ - $|3\rangle$ Feshbach resonance). The asymmetry in the trapping potential is $|\omega_x-\omega_y|/\omega_r \lesssim 0.01$. Note that $\omega_r \propto \sqrt{B}$ so the radial confinement also changes when we tune interactions. Typically, we produce clouds with temperatures of 0.09 $T/T_\mathrm{F}$, where $T_\mathrm{F}$ is the Fermi temperature and $\it{N}$ = 3$\times$10$^5$ atoms per spin state. 
\newline \indent To determine the temperature of the atom clouds (horizontal axis of Figs.~2 and 3 in the main text) we follow the quench procedure as described in the main text, however, instead of probing the clouds with Bragg lasers following the quench the cloud is held for 500~ms and an $in-situ$ absorption image is taken (after this hold time, no dynamics in the density distribution are detected~\cite{Dyke21}). The temperature can be determined by fitting the known equation of state (EoS) for the pressure of a unitary Fermi gas \cite{Ku12} to the line densities of a trapped atom cloud at equilibrium. Throughout this study we probe temperatures ranging from 0.1 to 0.18 $T/T_\mathrm{F}$. The higher temperatures are obtained by varying the endpoint of the initial evaporative cooling process, which loads a higher temperature cloud into the oblate trapping potential and the temperature is extracted following the same procedure described above.

\subsection{Bragg Spectroscopy Calibration}\label{sec:calibration}
\begin{figure}[t]
\includegraphics[width=0.5\columnwidth]{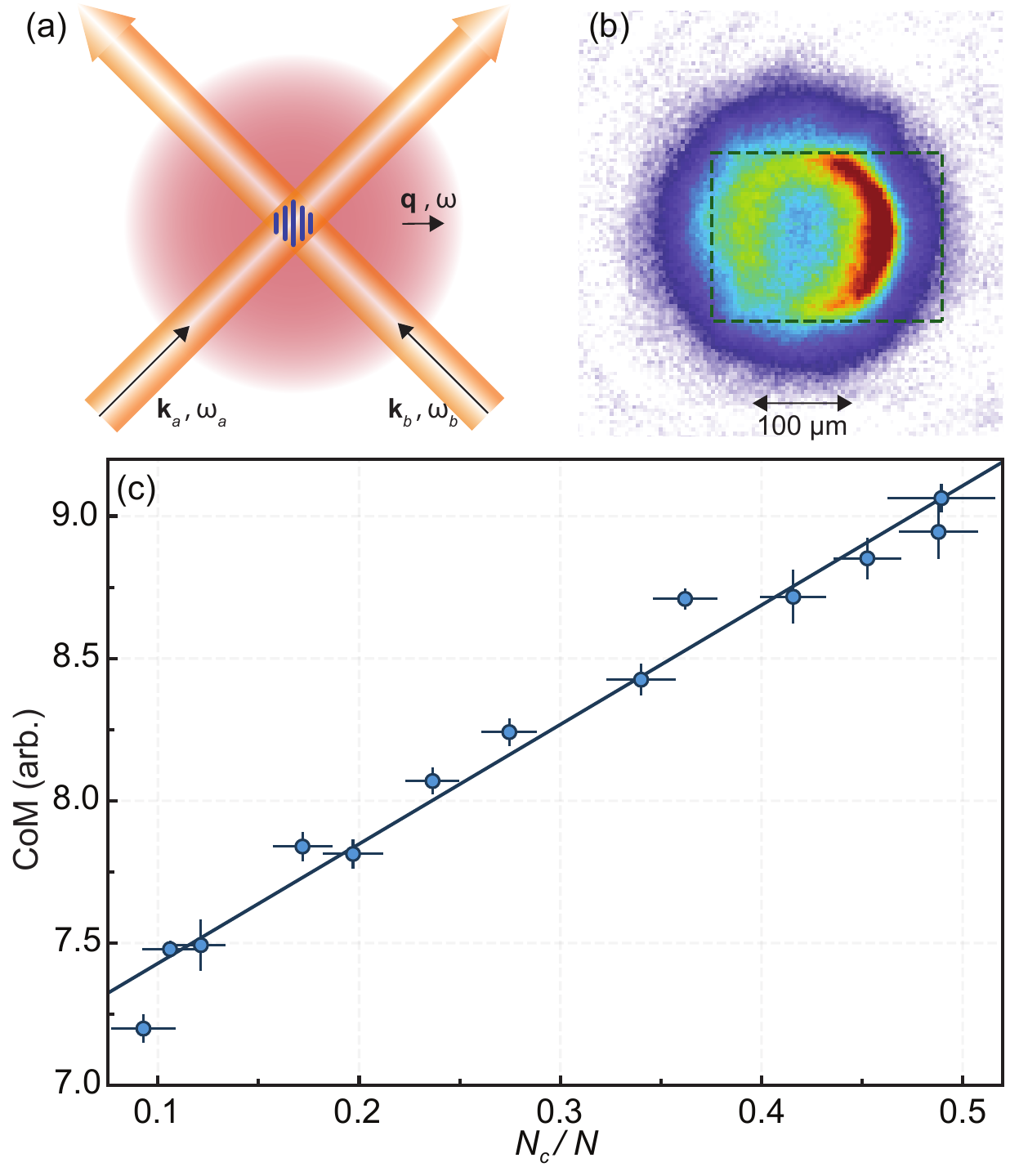}
\caption{(Color online) (a) Schematic of the experimental setup for focused beam Bragg spectroscopy. Two laser beams with wave vectors $\kk_a$ and $\kk_b$ and frequencies $\omega_a$ and $\omega_b$ are focused into the center of a trapped atom cloud. The beams have a 1/e$^2$ radius of 15 $\mu$m and intersect at an angle of 2$\theta$ =  72.6$^{\circ}$ (b) Absorption image of an atom cloud following an excitation with a Bragg frequency $\omega/(2 \pi)=+50$ kHz. The dashed rectangle indicates the region used for determination of the centre of mass displacement. (c) Measured center of mass displacement, $\Delta X_{\mathrm{CoM}}$ as a function of measured pair condensate fraction.}
\label{fig:SM1}
\end{figure}

The post-quench dynamics are probed using two tightly-focused Bragg laser beams, as shown in Fig.~\ref{fig:SM1}(a), that intersect in the centre of the cloud where the density is near-uniform~\cite{Hoinka17,Carcy19}. 
Two-photon Bragg spectroscopy is both energy and momentum selective thus, by setting the frequency difference to half of the atomic recoil frequency ($\omega_r/2=\hbar q^2/4m$), pairs with zero center-of-mass momentum are resonantly excited~\cite{Lingham14,Carcy19}. These pairs then begin moving with a velocity $\hbar \qq /m$ where, $\qq = \kk_a - \kk_b$ is the difference of the wave-vectors of the two Bragg lasers. We use $q\simeq 4 k_F$ to ensure that $\omega$ is large compared to $E_\mathrm{F}$, and the pulse duration (50 $\mu $s) is relatively short with respect to the dynamics. The center of mass displacement $\Delta X_{\mathrm{CoM}}$ following time-of-flight expansion, reveals the momentum imparted to the atoms by the Bragg lasers. Fig.~\ref{fig:SM1}(b) shows an absorption image of an atom cloud following the experimental sequence described in the main text. A large number of atoms are seen to be displaced from the centre of the cloud towards the right of the image. The crescent shape in Fig.~\ref{fig:SM1}(b) is a result of collisions between scattered atoms and the unperturbed atoms in the cloud which occur following the Bragg pulse~\cite{Veeravalli08}.

We have measured the sensitivity of the centre of mass displacement, $\Delta X_{\mathrm{CoM}}$  at unitarity as the pair condensate fraction varies. To determine the centre of mass displacement, $\Delta X_{\mathrm{CoM}}$, we prepare clouds at a range of initial temperatures by varying the endpoint of the evaporation and then applying a Bragg pulse (to the clouds at equilibrium) and measure the response as described above. 
To determine the pair condensate fraction a degenerate cloud at 832.2 G is produced at a given temperature and in equilibrium. Subsequently, we jump the magnetic field far onto the BEC side of the Feshbach resonance in $\approx$ 50 $\mu $s and simultaneously turn off the optical ($z$) confinement. The weakly bound pairs are converted to tightly bound molecules, whose centre of mass momentum is preserved~\cite{Regal04}. The weakly interacting molecules then expand ballistically along $z$ before the magnetic field is ramped back to 832.2~G in 2.5~ms to dissociate the molecules. The total expansion time is approximately one quarter of the radial trapping period, to best reveal the momentum distribution. Finally, an absorption image is taken and fitted with a bi-modal distribution to determine the condensate fraction. 

Fig.~\ref{fig:SM1}(c) shows the response of the $\Delta X_{\mathrm{CoM}}$ of the cloud as the pair condensate fraction, and therefore the temperature, is varied. As the condensate fraction is reduced, the centre of mass displacement, $\Delta X_{\mathrm{CoM}}$ also decreases. The Bragg signal shows a clear dependence on the condensate fraction, which reflects the change in the number of condensed pairs available to take part in the scattering process.

\subsection{Determination of Density}\label{sec:detdens}
To observe the Higgs oscillation most clearly the Bragg scheme addresses atoms at the centre of the trap, where the density is near homogeneous, as in previous studies~\cite{Hoinka17,Carcy19}. We define the mean density $\bar{n}$ in the Bragg volume as 
\begin{equation}
\bar{n} = \frac{\int \Omega_{\mathrm{Br}}(\rr) n(\rr) \mathrm{d}^3 \rr}{\int \Omega_{\mathrm{Br}}(\rr) \mathrm{d}^3 \rr},
\label{eq:nbar_exp}
\end{equation}
where $\Omega_{\mathrm{Br}}(\rr)$ is the spatially dependent two-photon Rabi frequency, which is proportional to the geometric mean of the intensities of the two laser beams $\sqrt{I_a(\mathbf{r})I_b(\mathbf{r})}$ and $n(\rr)$ is the 3D density distribution.
The mean density sets the relevant energy scale and the Higgs oscillation frequency via Fermi energy $E\mathrm{_F} = \frac{\hbar}{2m}(3\pi^2\bar{n})^{2/3}$, and the normalized frequency $\hbar\omega_{\rm H} / 2 E\mathrm{_F}$, where $m$ is the atomic mass and $\omega_{\rm H}$ is the Higgs oscillation frequency. 

\begin{figure}[t]
\centering
\includegraphics[width=0.8\columnwidth]{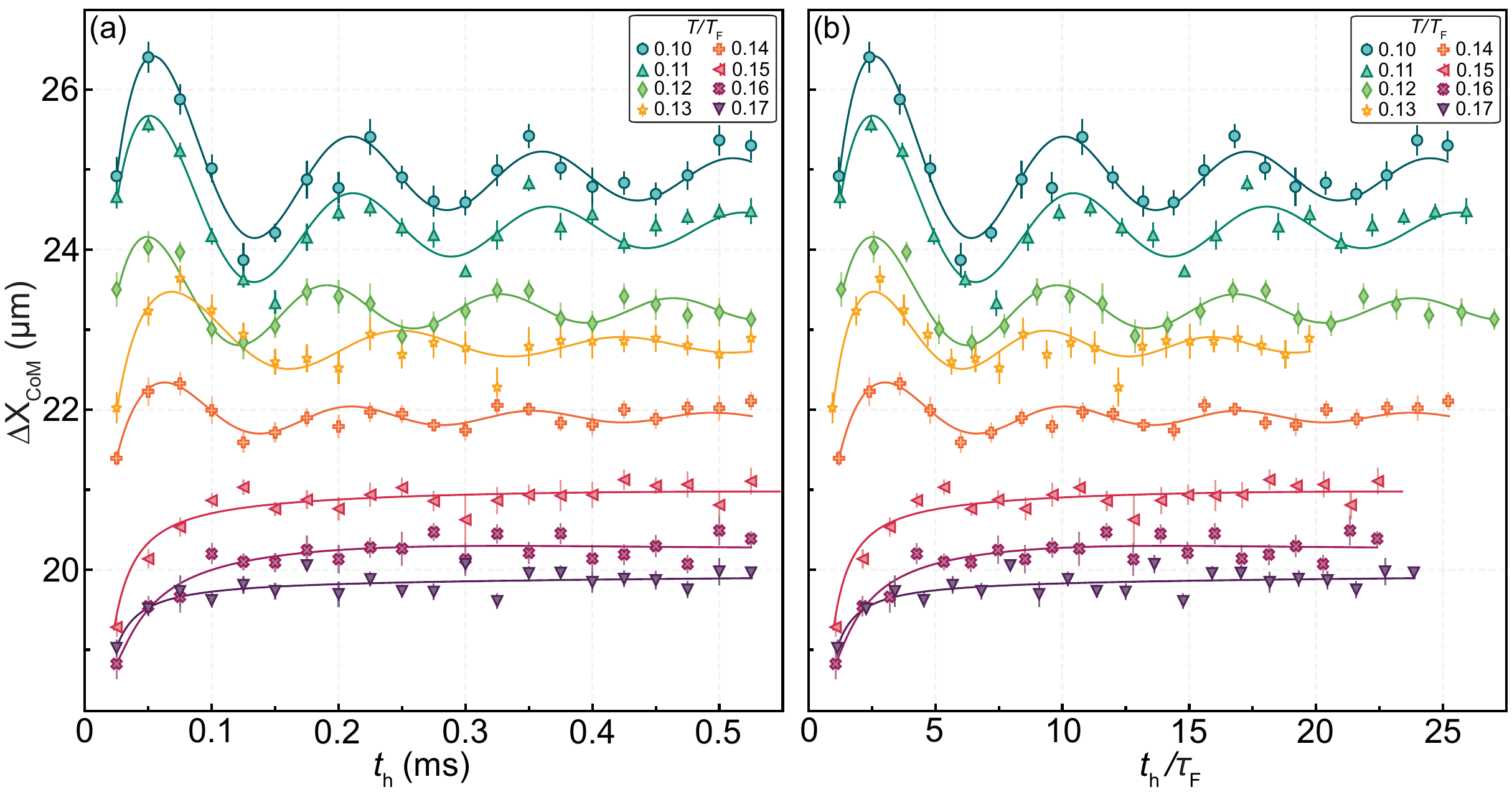}
\caption{(Color online) A selection of measured centre of mass displacement ($\Delta X_{\mathrm{CoM}}$) vs hold time, $t_h$ in absolute units of $\mu$s in (a) and relative to the local Fermi time $\tau_F = \hbar/E_F$ in (b) for atom clouds prepared at different temperatures. Also plotted are the fitted sinusoidal functions used to determine the frequency and damping of the amplitude oscillation. Different temperature clouds can have different densities leading to the different absolute oscillation frequencies, (a). When the time axis is taken relative to the Fermi time, (b) the relative frequencies agree more closely, as seen in the main text.}
\label{fig:SM_osc}
\end{figure}

To determine the mean density, the true (3D) density distribution $n$($\mathbf{r}$) of the trapped cloud and the intensity product of the Bragg laser beams with their intersecting Gaussian spatial profiles are combined. The 3D density $n$($\mathbf{r}$) is extracted through applying an inverse Abel transform to absorption images of unperturbed trapped clouds. Images of trapped clouds are taken using a high intensity ($I/I_{\mathrm{sat}} > 10$) and short imaging pulse length of 1~$\mu s$. The inverse Abel transform method uses a Fourier decomposition where the radial density distribution is expanded in a Fourier series. This method requires no direct differentiation and allows reconstruction of the central density without singularities. From this, we are able to determine $\bar{n}$ with an error of approximately 7.5~\%, which leads to a 5~\% error in the Fermi energy.

In Fig.~\ref{fig:SM_osc}, we plot the direct measured centre of mass displacement ($\Delta X_{\mathrm{CoM}}$) vs.~time along with the fitted sinusoid, for the same data as in Fig.~2 of the main text. These traces show how the total Bragg signal decreases with increasing temperature, the magnitude of the oscillations falls off with increasing temperatures and the frequency changes for clouds prepared at different temperatures. In Fig~\ref{fig:SM_osc}(a) the hold time is in ms and in \ref{fig:SM_osc}(b) in units of the local Fermi time $\tau_F = \hbar/E_F$. The absolute oscillation frequency is set by the local Fermi energy, which in turn is set by the local density. Clouds at different temperatures may have different densities due to changes in the preparation sequence. This leads to the variations in the absolute oscillation frequency in (a), which become more consistent when plotted as a function of the scaled time in (b).

\section{Order parameter dynamics within time-dependent BCS theory}\label{sec:numerics}
We recall here what is theoretically expected for the time evolution of a superfluid Fermi gas after a rapid change of the interatomic interactions at zero temperatures.  We consider a two-component ($\up$ and $\down$) Fermi gas interacting through a pairwise single-channel $s$-wave interaction, which captures broad, entrance-channel dominated Feshbach resonances~\cite{art:chin_feshbach}. Assuming the gas is homogeneous in a cubic volume $V$ (see Sec.~\ref{sec:exp} for consideration of inhomogeneities in the trapped gas) the momentum representation of the Hamiltonian for this system is 
\begin{equation}
\hat{H}= \sum_{\kg\sigma} \frac{k^2}{2m} \had_{\kg\sigma} \ha_{\kg\sigma} + \frac{g_0}{V} \sum_{\kg, \kg\aprime, \qg} \had_{\kg+\qg, \up}\had_{-\kg, \down} \ha_{-\kg', \down}\ha_{\kg'+\qg, \up},
 \label{eq:H_k}
\end{equation}
where the interaction strength $g_0$ is 
renormalized to reproduce the correct $s$-wave scattering length $a$ of the two-body problem.  In the remaining sections, we use the convention $\hbar=1$ throughout.

The equations of motion for the momentum distribution $n_\kg$ and pairing function $c_\kg$  (Eqs.~(1) and (2) in the main text) are derived from the Hamiltonian in Eq.~\eqref{eq:H_k},
using the BCS mean-field approximation. At $t=0$ the system is at equilibrium at scattering length $a_i$ and temperature $T<T_{c, i}$, we can therefore fix the initial conditions for $n_\kg$ and $c_\kg$ using the BCS ground-state solutions, as discussed in the main text
\begin{eqnarray}
n_\kg(t=0) &=&  \frac{1}{2} \left(1 -\frac{\xi_{\kg, i}}{\epsilon_{\kg, i}} F_\beta(\epsilon_{\kg, i})\right),\label{eq:nk0}\\
c_\kg (t=0) &=& - \frac{\Delta_{i}}{2\epsilon_{\kg, i} } F_\beta(\epsilon_{\kg, i}),\label{eq:ck0}
\end{eqnarray}
where $\xi_{\kg, i} = k^2/2m - \mu_i$, $\epsilon_{\kg, i} =\sqrt{\xi_{\kg, i}^2 + \Delta_i^2}$, $\mu_i$ and $\Delta_{i}$ are calculated at the initial scattering length $a_i$, and where $F_\beta(\epsilon)= \tanh(\beta \epsilon/2)=1-2(\mathrm{exp}(\beta \epsilon)+1)^{-1}$
is the thermal distribution with $\beta=1/k_\mathrm{B}T$. The limit of zero temperature corresponds to $F_\beta(\epsilon)\to 1$ for $\epsilon>0$.
Conversely, the regime of temperatures close to the critical temperature $T_c$ is found by taking the limit $\Delta\to 0$~\cite{art:review_BCSBEC}.

\subsection{Analytical solution in the small-amplitude regime}\label{sec:linear}
For shallow quenches, that is, small deviations of $n_\kk$ and $c_\kk$ from their equilibrium values,
the time-dependent BCS equations (Eqs.~(1) and (2) of the main text) can be solved analytically. At zero-temperature, this is a well-studied problem \cite{art:volkov1973,Gurarie2009,Foster2015}. Here we only detail the expression of the normalized amplitude $A_{\rm th}$ of the oscillations (see Eq.~(4) of the main text).

In the final equilibrium state (at $a=a_f$ and $T=0$) and in the modulus-phase basis \cite{higgs}, the linear-response matrix of the order-parameter is given by
\be
M(\omega)=
\begin{pmatrix}
\omega^2f(\omega) &  \omega g(\omega) \\
\omega  g(\omega) & (\omega^2-4\Delta_f^2)f(\omega)
\end{pmatrix}, \label{M}
\ee
where $\Delta_f$ and $\mu_f$ are the gap and chemical potential, respectively, of the final equilibrium state,
and we have introduced the functions:
\bea
f(\omega)&=&\frac{\Delta_f}{V}\sum_\kk\frac{1}{2\epsilon_\kk(\omega^2-4\epsilon_\kk^2)}, \\
g(\omega)&=&\frac{\Delta_f}{V} \sum_\kk\frac{ E_\kk-\mu_f }{\epsilon_\kk(\omega^2-4\epsilon_\kk^2)}.
\eea

The amplitude-amplitude response function introduced in Eq.~(3) of the main text follows directly from $M$ through the relation
\be
\chi_{|\Delta||\Delta|}(\omega)=(M^{-1}(\omega))_{22}=\frac{f(\omega)}{(\omega^2-4\Delta_f^2)f^2(\omega)-g^2(\omega)}.
\ee
As mentioned in the main text, the imaginary part $\chi_{|\Delta||\Delta|}''$ of this function has a squareroot divergence in $\omega=2\Delta$ for $\mu_f>0$ and
a square root cancellation for $\mu_f<0$. We denote by $f_l$ the spectral weight of this squareroot singularity:
\be
\chi_{|\Delta||\Delta|}''(\omega)\underset{\omega\to\omega_{\rm th}}{\sim}
\begin{cases}
f_l\sqrt{\frac{\omega_{\rm th}}{\omega-\omega_{\rm th}}} \text{ when } \mu_f>0 \text{ and } \omega_{\rm th}=2\Delta_f \\
f_l\sqrt{\frac{\omega-\omega_{\rm th}}{\omega_{\rm th}}} \text{ when } \mu_f<0 \text{ and } \omega_{\rm th}=2\sqrt{\Delta_f^2+\mu_f^2}
\end{cases}.
\label{chilowfreq}
\ee

In general, linear response theory expresses time-dependent quantities as Laplace transforms
of frequency responses. For the evolution of the amplitude of the order parameter, with initial conditions
Eqs.~\eqref{eq:nk0} and \eqref{eq:ck0}, this gives
\be
\frac{\Delta(t)-\Delta_f}{\Delta_i-\Delta_f}=1-\frac{1}{\chi_{|\Delta||\Delta|}(0)}\int_{+\infty+\ii\eta}^{-\infty+\ii\eta}\frac{\eee^{-\ii zt}}{2\ii\pi z}\chi_{|\Delta||\Delta|}(z).
\label{contour1}
\ee
This integral is deformed to a contour enclosing the branch cut $[2\Delta,+\infty)$ of $\chi_{|\Delta||\Delta|}$, as well as the pole of the integrand in $z=0$. Since only the contribution
of the pole survives at long times, one sees that $\lim_{t\to+\infty}\Delta(t)-\Delta_f = 0$, in other words
\be
\Delta_{\infty}=\Delta_f
\ee
in the small-amplitude regime and at $T=0$. From the contribution of the branch cut
we then derive
the version of Eq.~(4) of the main text that is valid at all times:
\be
\frac{\Delta(t)-\Delta_f}{\Delta_i-\Delta_f}=\frac{2}{\pi } \int_{2\Delta_f}^{+\infty}\frac{\cos \omega t}{ \omega}\frac{\chi''_{|\Delta||\Delta|}(\omega)}{\chi_{|\Delta||\Delta|}(0)}\dd\omega.  \label{contour2}
\ee
At long times, this frequency integral is evaluated by approximating $\chi''_{|\Delta||\Delta|}$ by its behavior (Eq.~\eqref{chilowfreq}) near the pair-breaking threshold.
This yields the expression of $A_{\rm th}$ (see Eq.~(4) of the main text):
\be
A_{\rm th}\underset{T=0}{=} \frac{2f_l}{\sqrt{\pi} \chi_{|\Delta||\Delta|}(0)} 
\ee
with $\chi_{|\Delta||\Delta|}(0)$ the static modulus response and $f_l$ the
spectral weight of the pair-breaking threshold.
\subsection{Numerical simulations}

Outside the small amplitude regime, we evolve the equations of motion (Eqs.~(1) and (2) of the main text) using a Runge-Kutta method~\cite{press1989numerical}.
The momentum-space integrals in the gap ($\Delta=(g_0/V)\sum_\kk c_\kk$) and number ($N=2\sum_\kg n_\kg$) equations are evaluated in spherical coordinates using a fine grid in momentum space\footnote{Numerically, we evaluate the momentum-space integral of a generic function $f(k)$ using
 $\int d^3k/(2\pi)^3 f(k) \rightarrow \sum_i w_i k_i^2 f(k_i)/2\pi^2$, where $k_i$ and $w_i$ are the abscissas and weights of the Gauss-Legendre quadrature method~\cite{press1989numerical}. 
 Because $n_\kg$ and $c_\kg$ become small after a few $k_F$, we split the integration in two intervals: $[0, 10k_F]$ and $[10 k_F, \Lambda]$, where $\Lambda=350 k_F$ is a momentum cut-off (large enough such that numerical results
are insensitive to it). In every interval, we take $1500$ points, so that the first interval is much more dense than the second one, where $n_\kg$ and $c_\kg$ are $\approx 0$.}. The conservation of the total number of atoms $N$ provides a check on the integrity of the numerics. Similar dynamical problems near resonance have been studied in Refs.~\cite{PhysRevLett.87.120406, PhysRevA.65.053617}.

\begin{figure}
  \centering
\includegraphics[width=0.5\columnwidth]{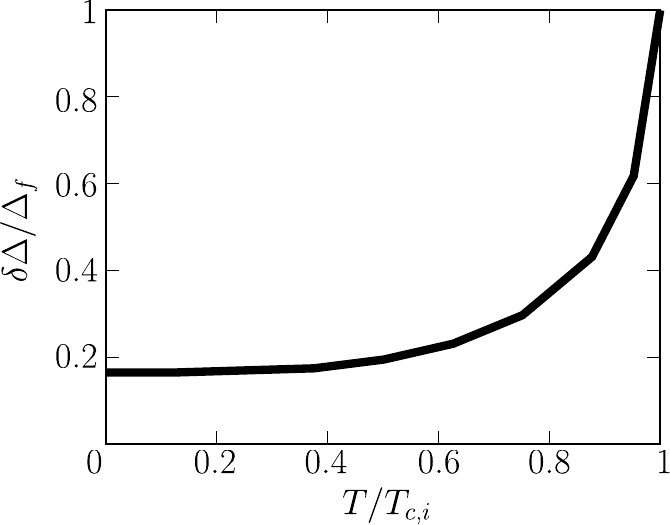}
\caption{ Temperature dependence of the difference in the equilibrium order parameter from the initial and final state $\delta\Delta \equiv |\Delta_{f} - \Delta_{i}|$. The initial and final state correspond to $1/(k_Fa_i)=-0.18$ and $1/(k_Fa_f)=0$, respectively. The temperature is rescaled by the critical temperature at $a_i$, that is $T_{c, i}=0.399 T_F$}.
\label{fig:DeltaD_T}
\end{figure}

In this section, we consider an infinitely fast quench from $1/(k_Fa_i)=-0.18$ to $1/(k_Fa_f)=0$ (see Sec.~\ref{sec:exp} for a discussion of the nonzero ramping time). As shown in Fig.~\ref{fig:DeltaD_T}, the relative variation of the equilibrium order parameter is small at $T=0$ ($\delta\Delta \equiv |\Delta_{f} - \Delta_{i}| \approx 0.16 \Delta_f$) but  increases with temperature, tending to 1 at $T_c$. Therefore, this limits the applicability of the analytical results presented in Sec.~\ref{sec:linear} to the small temperature regime.  

\begin{figure}
\centering
\includegraphics[width=0.7\columnwidth]{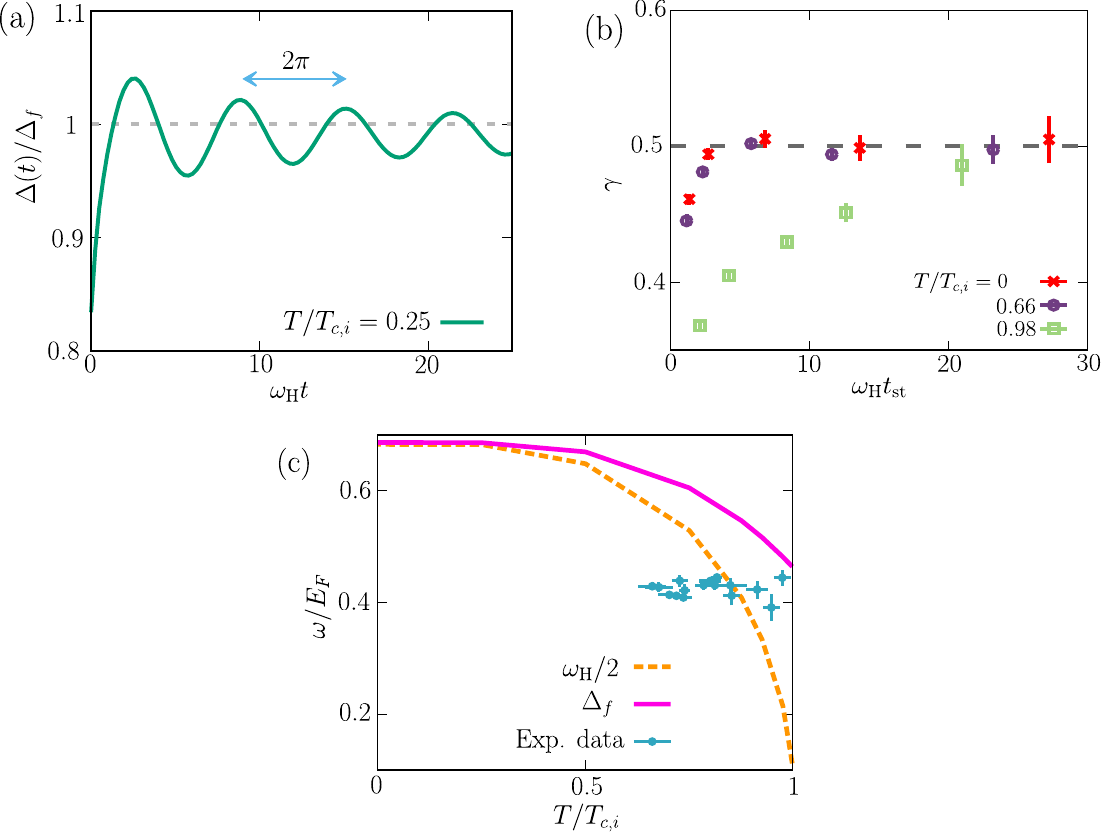}
\caption{(a) Numerical result for the time evolution of the amplitude of the order parameter $\Delta(t)$ at $T=0.25 T_{c,i}$ after a fast quench from $1/(k_Fa_i)=-0.18$ to $1/(k_Fa_f)=0$.  {The frequency of oscillation is $\omega_{\rm H}= 2 \Delta_\infty\lesssim  2 \Delta_f$.  (b) Damping coefficient $\gamma$ for $\Delta(t)$ as a function of the starting point of the fit $t_\mathrm{st}$ for different temperatures.  Here, we have used as a fitting function $\Delta(t)=\Delta_\infty+A\cos(\omega_{\rm H} t + \phi)/t^\gamma$ in the range $[t_\mathrm{st}, t_\mathrm{end}]$, where the final point $t_\mathrm{end}$ is fixed.} (c) Oscillation frequency (dashed dot line) as a function of temperature $T/T_{c, i}$.  The dashed (purple) line corresponds to $\Delta_f$ in Fermi units. The data points are from the experimental results in Fig.~3(a) of the main text.}
\label{fig:Delta_time}
\end{figure}

Using numerical simulations, we can instead explore the entire temperature range $ 0 \leq T \lesssim T_{c,i}$.
For example, Fig.~\ref{fig:Delta_time}(a) shows the time evolution of the amplitude of the order parameter $\Delta$ at temperature $T=0.25 T_{c,i}$. As expected from literature (see e.g. Ref.~\cite{Foster2015} for $T=0$) and discussed in the main text, the order parameter asymptotes to a value $\Delta_\infty$ a bit less than the expected final state equilibrium value $\Delta_f$ and oscillates at a frequency $\omega_{\rm H} =2\Delta_\infty$. {Moreover, we expect a power-law damping coefficient $\gamma=0.5$, which is found if we neglect the very short time dynamics, as shown in Fig.~\ref{fig:Delta_time}(b) for different temperatures.} Finally, in Fig.~\ref{fig:Delta_time}(c), we show the temperature dependence of the Higgs oscillation frequency (orange dash-dotted line) and the final order parameter (pink solid), and compare it to the experimental data.
The temperature is scaled to the critical temperature of the initial scattering length (for the experimental data we have used  $T_{c, i} = 0.15 T_F$). The comparison is not meant to be quantitative, as BCS theory is known to overestimate both the superfluid gap and critical temperature at unitarity \cite{art:review_BCSBEC}, but rather to confront the trend in the vicinity of $T_c$.

\section{Application of time-dependent BCS theory to the experimental setup}\label{sec:exp}
In this Section, we consider the impact of various experimental effects on the amplitude oscillations and their detection by Bragg spectroscopy. This includes the following:  the presence of a nonzero-duration initial ramp of the magnetic field (Sec.~\ref{sec:ramp-rate}), density inhomogeneities due to the trapping potential (Sec.~\ref{sec:LDA}), and time averaging over a Bragg pulse of nonzero duration (Sec.~\ref{sec:BraggT}).
\subsection{Ramp-rate dependence}\label{sec:ramp-rate}
To mimic experimental protocols for Feshbach resonances, we include in our numerics an initial linear magnetic field ramp:  $B(t)=B_i+ R t$, where $R=dB/dt$ is the ramp rate.
The presence of the ramp delays the start of the Higgs oscillation. This delay can be quantified using a timescale $t_q=|B_f-B_i|/ R$, which must be compared with the Fermi scale $\tau_F$ and the characteristic timescale of the order parameter dynamics  
$t_\Delta= 1/\Delta$~\cite{art:Barankov2004, art:Scott}. When $t_q$ becomes much longer than $t_\Delta$, $|\Delta(t)|$ adiabatically evolves to  $\Delta_f$.

\begin{figure}
  \centering
\includegraphics[width=0.5\columnwidth]{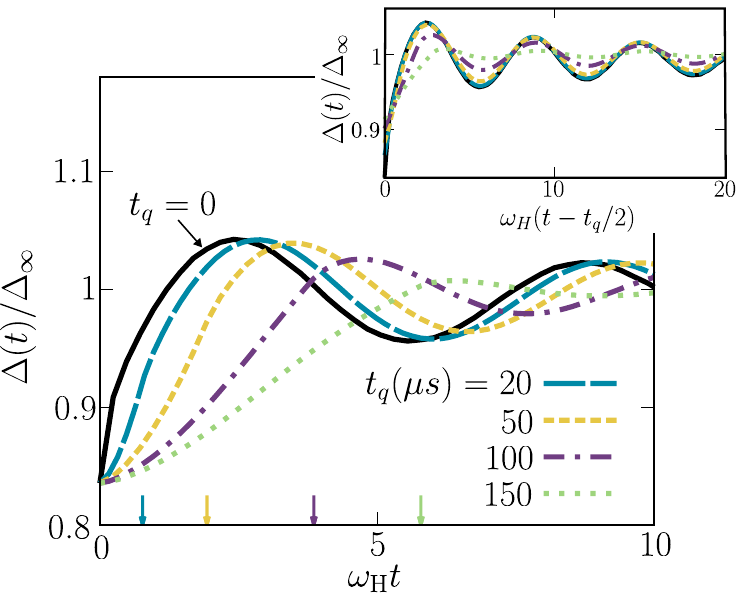}
  \caption{Early-time evolution of the amplitude of the order parameter $\Delta(t)$ after a quench from $1/(k_Fa_i)=-0.18$ to  $1/(k_Fa_f)=0$ for an infinitely fast ramp (black solid line) and for  different ramp time $t_q$ (coloured lines) at zero temperature. We indicate with the bottom coloured arrows, the values of the ramp time $t_q$ corresponding to ramp rate $R$ (see the text). Inset: Same as (a) but the curves are delayed by {half the ramp time $t_q/2$}.} 
  \label{fig:delta_R}
\end{figure}

 We vary the ramp rate $R$ to study its impact on the early-time dynamics of $|\Delta(t)|$. Fig.~\ref{fig:delta_R} shows how for slower changes of $B$ (smaller $R$) 
 the amplitude of the oscillations becomes smaller leading to a decrease in the visibility of the Higgs mode, and the early-time behaviour of $|\Delta|$ has a more parabolic shape that cannot 
 be fitted using a function $\sim\cos (2\Delta t)/\sqrt{t}$ (see Eq.~(4) in the main text). In the inset of Fig.~\ref{fig:delta_R}, we plot the same quantities 
 but we include a delay equal to half ramp time $t_q/2$. Accounting for this delay, acts to align the oscillation phases.

\subsection{Inhomogeneous broadening of the signal}\label{sec:LDA} 
As discussed in Sec.~\ref{sec:preparation}, during the sample preparation, the atoms are loaded into an oblate harmonic potential. In this section, we show how this causes an inhomogeneous broadening of the signal, which can lead to a sharp increase of the damping coefficient $\gamma$ from $1/2$ to 2 depending on the width of the Bragg beam. 

We begin by writing the external trapping potential as
\begin{equation}
V_\mathrm{ho}(\rg)= \frac{1}{2}m \left(\omega_r^2 r^2 + \omega_z^2 z^2 \right),
\label{eq:Vext}
\end{equation}
where $\omega_r$ and $\omega_z$  are the trapping angular frequencies in the radial and axial directions, respectively~\cite{pitaevskii2016bose}. 
In the local density approximation (LDA), the trapped system can be treated as a collection of locally  uniform subsystems\footnote{The local density approximation remains valid while the harmonic oscillator length, which is the typical length at which $n(\rg)$ varies, is much larger than both the healing length $1/\sqrt{2m\mu}$ and pair correlation length~\cite{Strinati1998} with the latter diverging near $T_c$ as $ k_F/m\Delta$, marking an obvious breakdown of the approximation.}. This means that at equilibrium the equation of state for the superfluid can be solved after replacing $\mu$ by a local $\mu(\rg)=\mu-V_\mathrm{ho}(\rg)$ 
and the number equation becomes
\begin{equation}
N = \int d^3r n(\rg),
\label{eq:N_rc}
\end{equation}
where $n(\rg)=2\sum_\kg n_\kg(\rg)$ is the local density. Consequently, $\Delta(\rg)$, $1/(k_F(\rg) a)$, and $k_\mrm{B}T/E_\mrm{F}(\rg)$ also acquire a dependence on $\rg$~\cite{art:Perali_2003, art:Salasnich_cond, art:Chiofalo_sign, art:Haussmann_2008}. {At unitarity, $1/(k_F(\rg) a)=+\infty$ becomes uniform, but $\Delta$ still depends on $\rr$ because $T/T_c(\rr)$ does.}

At zero temperature, for a  weakly interacting gas ($|a| \ll k_\mrm{F}^{-1}$), $n(\rg)$ is well-approximated by the {Thomas-Fermi distribution}:
\begin{equation}
 n^{(0)}(\rg)= \frac{8 N}{\pi^2 R_r^2 R_z} \left[1 - \left(\frac{r}{R_r}\right)^2-\left(\frac{z}{R_z}\right)^2 \right]^{3/2},
 \label{eq:nr_TF}
 \end{equation}
 where $R_i=\sqrt{2E_F^\mrm{ho}/(m\omega_i^2)}= a_\mathrm{ho} (24 N)^{1/6} \omega_\mathrm{ho}/\omega_i$  ($i=\{r,z\}$) are the Thomas Fermi radii, $\omega_\mathrm{ho}=(\omega_r^2 \omega_z)^{1/3}$ is the geometric mean of the angular frequencies, $a_\mathrm{ho}=\sqrt{1/(m\omega_\mathrm{ho})}=(24 N)^{1/6}/k_F^\mathrm{ho}$ is the oscillator length, and $k_F^\mrm{ho}=\sqrt{2mE_F^\mathrm{ho}}$ is the Fermi momentum in a trapped system. 
 
\begin{figure}
\centering
\includegraphics[width=0.35\columnwidth, angle=270]{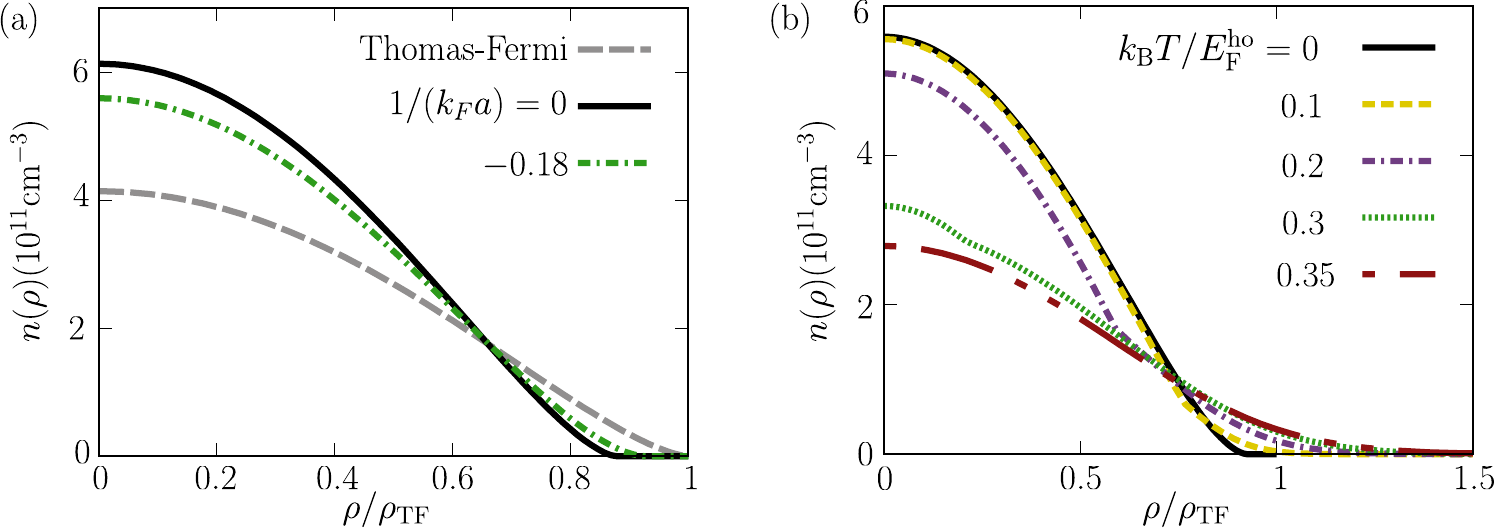}
\caption{Particle density $n(\rho)$ in a harmonic trap as a function of the weighted radial coordinate  $\rho$ in Fermi units.
(a) At zero temperature for a non-interacting system (dashed grey), for $1/k_\mrm{F}^\mrm{ho}a=-0.18$ (green dash dotted), and at unitarity (black solid). (b) For fixed $1/(k_\mrm{F}^\mrm{ho}a)=-0.18$ and different temperatures (different colors). }
\label{fig:nR}
\end{figure}

Equation~\ref{eq:nr_TF} is altered by increasing the scattering length with respect to $k_\mrm{F}^\mrm{ho}$ as shown in Fig.~\ref{fig:nR}(a), where we compare $n^{(0)}({\bf r})$ with the density profiles for $1/(k_F^\mathrm{ho} a)=-0.18$ and $0$ using the trapping parameters of Sec.~\ref{sec:preparation}\footnote{When $a\to \infty$, the dimensionless parameter $1/(k_F(\rg) a)$  loses spatial dependence, and the equation of state for the Fermi gas is expected to be universal in terms of density scales and is written in terms of the Bertsch parameter $\xi_\mathrm{B}$~\cite{art:Ho, art:review_BCSBEC}. Therefore, using $\mu(\rg)/E_F(\rg)=\xi_\mathrm{B}$ and Eq.~\ref{eq:nr_TF}, the density profile can be approximated by $n(\rg)=\xi_\mathrm{B}^{-3/2}n^{(0)}(\rg)$, which well describes experimental data in the unitary regime when a beyond-mean field $\xi_\mathrm{B}$ is used~\cite{art:Perali_n}.}.
In Fig.~\ref{fig:nR}, we have utilized the weighted radial coordinate 
\begin{equation}
\rho=\frac{\sqrt{(\omega_r r)^2+ (\omega_z z)^2}}{\omega_\mathrm{ho}},
\label{eq:rho}
\end{equation}
 with corresponding (rescaled) Thomas-Fermi readius $\rho_\mathrm{TF}=\sqrt{2 E_\mrm{F}^\mrm{ho}/(m\omega_\mathrm{ho}^2)}$. 
From Fig.~\ref{fig:nR}(a), it is clear that interactions lead to a narrowing of the density distribution. It is noteable that Eq.~\eqref{eq:rho} also shows how the case of a cigar-shape trap can be reduced to the spherical case by homothety. 
The influence of temperature on the density profile is shown in Fig.~\ref{fig:nR}(b). Deviations from the zero-temperature profile (black solid curve) begin on the edges of the cloud and progress towards the center as temperature is increased.

Within the local density approximation, we use the density distribution $n(\rg)$ to calculate the global order parameter by averaging over the cloud profile. The spatial weight is the product of the local density, $n(\rg)$, and local intensity of the probe beam, $I(\rg)$,
\begin{equation}
{\bar\Delta(t)} = \frac{1}{\bar{n}}\int d^3r n(\rg)I(\rg) \Delta(\rg),
\label{eq:D_av_red}
\end{equation}
such that 
\begin{equation}
\frac{1}{\bar{n}}\int d^3r n(\rg) I(\rg) = 1,
\label{eq:nred_norm}
\end{equation} 
where $\bar{n}$ is the mean density.  The finite width of the probe beam allows us to consider smaller volumes of the cloud, mimicking the experimental Bragg scheme described in Sec.~\ref{sec:detdens}. We approximate the probe beam with a Gaussian profile, such that 
\begin{equation}
I(\rg)= I(0)\,\exp\left(-\frac{|\rg|^2}{2\lambda^2}\right),
\label{eq:I_Gau}
\end{equation}
where $I(0)$ is a normalization constant and $\lambda$ is the effective width of the beam which is fixed to match the experimental mean density (Eq.~\ref{eq:nbar_exp}).

Using Eq.~(4) of the main text for the time-evolution of $\Delta$, Eq.~\eqref{eq:D_av_red} becomes
\begin{equation}
{\bar\Delta(t)}=\frac{1}{\bar{n}}\int d^3r n(\rg) I(\rg)\left(\Delta_{\infty}(\rg)+\frac{\cos(2\Delta(\rg)t+\phi)}{\sqrt{\nu(\Delta(\rg)) t}}\right).
\label{eq:Delta_Gur}
\end{equation}
with the rewriting of the oscillation amplitude $\nu=2\Delta/(A_{\rm th}(\Delta_i-\Delta_\infty))^2$.
The density average of functions oscillating at frequencies $2\Delta(\rg)$ leads to a blurring of the Higgs
signal and consequently to a reduced contrast of the oscillations. 

\subsubsection{Qualitative discussion in the limit of a tightly focused beam} When the beam is well focused around the center of the trap, one can approximate the density using
\begin{equation}
n(\rho)=n_0-\alpha \rho^2,
\label{eq:nr_0}
\end{equation}
with $\alpha=-(1/2)d^2n/d\rho^2$ (see Eq.~\eqref{eq:nr_TF}). The small variation of the density around trap center causes a spatial variation of the order parameter $\Delta(\rho)=\Delta_0- \Delta' \alpha \rho^2$, and of the magnitude of the Higgs oscillations $\nu(\rho)=\nu_0-\nu'\alpha \rho^2$ (where
$\Delta'=d\Delta/d n$ and $\nu'=d\nu/d n$). To simplify the discussion, we consider here only a probe which respects the cylindrical symmetry of the trapping potential $I(\rg)=I(\rho)$, as defined in Eq.~\eqref{eq:I_Gau}. 
This leads to
\begin{equation}
{\bar\Delta(t)}={{\bar\Delta_{\infty}}}+\frac{4\pi  n_0}{\bar{n}\sqrt{\nu_0 t}} \int_0^{+\infty} \rho^2 d\rho I(\rho) (1-c \rho^2)  \cos(2\Delta_0 t -\Delta' \alpha \rho^2t+\phi),\label{deltacylinder}
\end{equation}
where the term proportional to $c=\alpha(1/n_0-\nu'/2\nu_0)$ is included for completeness
but becomes negligible when the beam waist tends to 0. In this approximation, the averages of slowly-varying quantities
simply coincide with their value at the center of the trap ($\bar{n}=n_0$, $\bar{\nu}=\nu_0$ etc.),
and only the average of the oscillatory part remains. 

Performing the radial integral in Eq.~\eqref{deltacylinder} for the Gaussian profile Eq.~\eqref{eq:I_Gau} gives
\begin{equation}
{\bar\Delta(t)}={{\bar\Delta_{\infty}}}+\frac{1}{\sqrt{\nu_0 t}} \text{Re} \left[\eee^{2\ii\Delta_0t+\ii\phi} f(t)\right],
\label{avDelta}
\end{equation}
where the complex-valued blurring function
\begin{equation}
f(t)=\left(\frac{1-\ii t/t_{\rm d}}{1+t^2/t_{\mathrm d}^2}\right)^{3/2}
\end{equation}
depends on the characteristic timescale
\begin{equation}
t_{\rm d}=\frac{1}{4\lambda^2 \alpha \Delta'}\approx\frac{1}{\Delta(\lambda)-\Delta_0}
\label{eq:td}
\end{equation}
Assuming that this timescale is large ($\Delta_0 t_{\rm d}\gg 1$), 
the Higgs oscillations will still display the expected $1/\sqrt{t}$ attenuation law at short times ($t\ll t_{\rm d}$).
However, beginning at $t\approx t_{\rm d}$ the decay is quickened and the phase shifts from the original $\phi=\pi/4$ phase.  At later times ($t\gg t_{\rm d}$), one finds instead a $1/t^2$ attenuation law:
\begin{equation}
{\bar\Delta(t)}\underset{t\to+\infty}{\simeq}{{\bar\Delta_{\infty}}}-\frac{t_{\rm d}^{3/2}}{\sqrt{\nu_0}t^2}\sin 2\Delta_0 t.
\label{eq:deltat_tinf}
\end{equation}
This shows that inhomogeneous broadening can seriously impair the observation of the power-law damping exponent $\gamma=1/2$ or $3/2$.

\subsubsection{Numerical solutions in the general case}
\begin{figure}
\centering
\includegraphics[width=0.5\columnwidth]{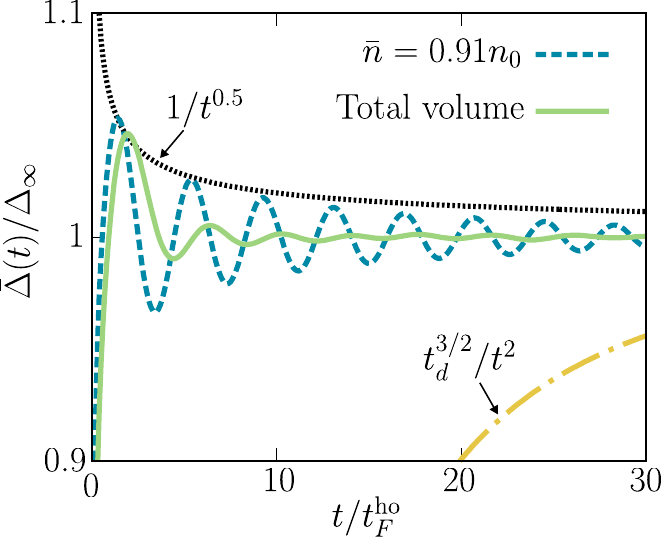}
\caption{Time evolution of the order parameter after averaging over the total density profile of the trap (green solid line) and over a reduced volume with corresponding $\bar{n}=0.91 n_0$ (blue dashed line) at zero temperature. Black dotted and yellow dashed dotted lines indicate the power law curves $1/\sqrt{t}$, which is found at short times, and $1/t^2$ from Eq.~\ref{eq:deltat_tinf}, which characterizes longer times than the maximum time in the figure ($t \gg t_\mrm{d}$).}
\label{fig:AvDelta_15}
\end{figure}

More generally, we solve numerically\footnote{
To perform the integration in Eq.~\ref{eq:D_av_red}, we sample $n(\rg)$ using a \qte{accept-reject method}~\cite{book:krauth}, which generates a set of random numbers distributed according to the initial distribution: $n(\rg)\rightarrow \{ n_i,$ with $i=1,\cdots, N_\mrm{p} \}$. The average on Eq.~\ref{eq:D_av_red} can be then performed using
\begin{equation}
 \bar{\Delta}(t) = \frac{1}{\sum_i^{N_\mathrm{p}} I(n_i)}\sum_{i=1}^{N_\mathrm{p}}\,I(n_i)\Delta (n_i, t),
 \label{eq:dt_av_samp}
 \end{equation}
  where $N_\mathrm{p}$ is the total of number of extracted values $n_i$. 
%
%
%
 } 
 Eq.~\eqref{eq:D_av_red} for different values of $\lambda$ and, consequently, for different $\bar{n}$.  Fig.~\ref{fig:AvDelta_15} shows the evolution of the superfluid order parameter at zero temperature after a quench from $1/k_F^\mrm{ho}a=-0.18$ to unitarity after averaging over a probe beam of width  $\lambda=0.06\rho_{\rm TF}$ such that $\bar{n}= 0.91 n_0$, compared to the case where the probe beam covers the entire volume. We observe that this latter case yields strongly-damped oscillations, which would be difficult to observe. Instead, the experimentally-relevant average over a smaller volume leads to distinct oscillations.
Moreover, because  $\lambda = 0.06\rho_{\rm TF}$  is small compared to the effective Thomas-Fermi radius $\rho_\mrm{TF}$, the characteristic time $t_d$ is much larger than  the oscillation period $2\pi/\omega_\mrm{H}$, and therefore, for times $t< t_{\rm d}$,  we do not expect the attenuation law $1/t^2$ (as described by Eq.~\eqref{eq:deltat_tinf}) to play a role within the experimentally relevant time window.

\begin{figure}
\centering
\includegraphics[width=0.5\columnwidth]{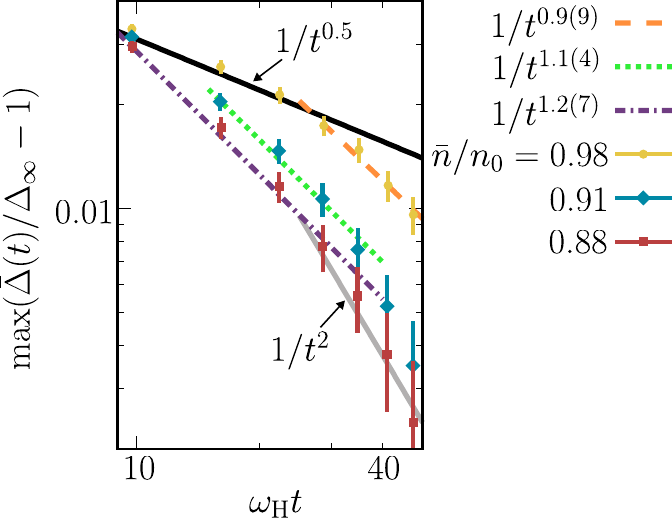}
\caption{Data points: Maximum values of $(\bar{\Delta}(t)/\Delta_\infty - 1)$ for different $\bar{n}$ or equivalently $\lambda$ at $T=0.66 T_\mrm{c, i}$. The solid black line represents the homogeneous power-law behavior $1/\sqrt{t}$ (see Eq.~(4) in the main text). Dashed lines: Fitting functions $1/t^\beta$ for times $t< t_d$. Gray solid line: power law $1/t^2$ recovered only in the case with lower $\bar{n}$.} 
\label{fig:peaks}
\end{figure}

We address also the impact of inhomogeneous broadening on the damping law in the time region $0< t < t_\mrm{d}$. Fig.~\ref{fig:peaks} shows trends in the local oscillation maxima $\bar{\Delta}(t_n)$ at discrete times $t_n$ measured in units of the oscillation frequency $\omega_{\rm H}$ and at temperature $T=0.66 T_\mrm{c, i}$. We fit the first few oscillations to a power-law damping, thereby
mimicking the analysis performed on the experimental data.
For smaller values of $\lambda$ compared to 
$\rho_\mrm{TF}$,  the power law decay $1/\sqrt{t}$ is still a good fit for the short-time behavior of $\bar{\Delta}$, as shown by the case with $\bar{n}=0.98 n_0$ in Fig.~\ref{fig:peaks}. For larger values of $\lambda$, we find an intermediate behavior, characterized by a power-law decay $1/t^\beta$ with $0.5 < \beta < 2.0$.

\subsection{Averaging over the Bragg time}\label{sec:BraggT}
\begin{figure}
\centering
\includegraphics[width=0.5\columnwidth]{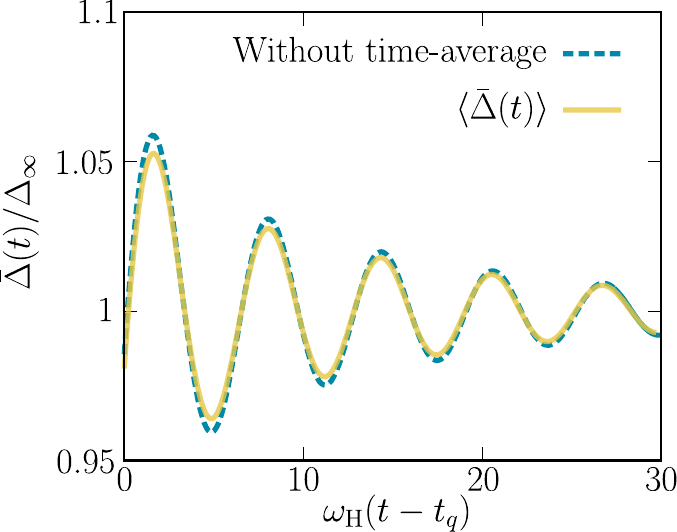}
\caption{Comparison between the time and space averaged order parameter ($\langle \bar{\Delta}(t) \rangle$, yellow solid line) and the same quantity only averaged over space ($\bar{\Delta}(t)$, blue dashed line). The time average is performed over the Bragg time $t_\mrm{B}$ (see Eq.~\eqref{eq:avt}) and the density average corresponds to the case $\bar{n} =0.91 n_0$ (see Eq.~\eqref{eq:D_av_red}). The temperature is  $T=0.66 T_\mrm{c, i}$.}
\label{fig:Delta_tBr}
\end{figure}

Experimental measurements are taken using two-photon Bragg scattering as illustrated in Fig.~1 of the main text. To avoid
over-broadening the sharp features of the response function, the duration $t_\mrm{B}=50 \mu s$
of the Bragg pulse should be as long as possible. However, in an out-of-equilibrium system, using pulses that are too long allows the system to evolve significantly during the probe time violating the timescale separation assumptions underlying Eq.~(5) of the main text and possibly blurring the oscillatory signal.
Explicitly, the observable $\Delta$ reported at the time $t$ is actually the result of a time-average: 
\begin{equation}
\braket{\Delta(t)}_\mrm{Bragg} = \frac{1}{t_\mrm{B}}\int_{t-t_\mrm{B}/2}^{t+t_\mrm{B}/2} \Delta (t') dt'.
\label{eq:avt}
\end{equation}
Fig.~\ref{fig:Delta_tBr} shows the comparison between the time-evolution of (spatially-averaged) $\Delta$ for $t_B=0$ and $t_B=50 \mu s$. 
Although the time-averaging reduces the contrast of the oscillations by reducing their the magnitude, it is less critical
than the spatial averaging as it does not affect the attenuation law.

\section{Comparing theoretical predictions and experimental measurements}\label{sec:comp}

{The comparisons with experiment shown in Figs.~2(b) and 3(b) of the main text use theoretical predictions for $\braket{\bar \Delta(t)}$ that combine both the spatial and temporal averages discussed in Secs.~\ref{sec:LDA} and \ref{sec:BraggT} and the dephasing introduced by the finite ramp time $t_q$ discussed in Sec.~\ref{sec:ramp-rate}.  From Eq.~(5) of the main text, we write the experimental signal as ${ S(t)} \approx \alpha + \beta { \Delta(t)}$, where  $\alpha$ and $\beta$ depend on the sensitivity with an explicit expression of those quantities given below. This dependency is removed by taking the ratio
\begin{equation}
\frac{{ S(t)}-S_\infty}{S_f - S_i} \sim \frac{\braket{\bar \Delta(t)}-\Delta_\infty}{\Delta_f - \Delta_i}.
\label{eq:Scomp}
\end{equation}
To avoid the well-known overestimation of BCS theory of the critical temperature $T_c$ and superfluid gap $\Delta$, we compare theoretical predictions and experiment signal as a function of $\omega_{\rm H}t$ and $T/T_{c,i}$ using experimental values of $\omega_{\rm H}/\epsilon_F$, fitted from the bare experimental signal, and $T_{c,i}/T_F$, estimated as the temperature at which Higgs oscillations are no longer observed.

%
%

\subsection{Comparisons including the sensitivity of the Bragg pulse}\label{sec:braggsens}

In this subsection, we predict the experimental Bragg signal from our calculation 
of the density-density response within the Random Phase Approximation (RPA), that is, without scaling out the sensitivity. 
As discussed in the main text (see Eq.~(5)), the Bragg signal is sensitive
to the variations of $\Delta(t)$ through the dimensionless sensitivity
\be
\sigma=\frac{\Delta}{\chi_{nn}}\bb{\frac{\partial\chi_{nn}}{\partial\Delta}}_{n,T},
\label{sigma}
\ee
This thermodynamic quantity is computed by comparing equilibrium states having the same equilibrium density and temperature
but slightly different values of the order parameter $\Delta$ 
(or equivalently, slightly different values of the scattering length $a$). The calculation of $\chi_{nn}$ within RPA
is a standard problem \cite{Mottelson2001,Castin2001}. In practice we have used the explicit expressions of Ref.~\cite{repdens}
(see in particular Eqs.~(36) and (46) therein) to evaluate numerically $\chi_{nn}$ as a function of the excitation frequency $\omega$ and wavevector $q$.

Fig.~1b of the main text shows $\chi_{nn}(\omega)$ in the regime of large excitation wavevector ($q=4k_F$) used in the experiment.
The feature which characterizes the superfluid phase is a sharp edge at the pair-breaking threshold:
\be
\omega_{\rm th}=
2\sqrt{\Delta^2+\bb{\frac{q^2}{8m}-\mu}^2} \quad \text{ (for } {q^2}/{8m}>\mu).
\label{omegath}
\ee
As this threshold location changes with $\Delta$, the density response $\chi_{nn}$ for fixed
$\omega/\epsilon_F$ varies sharply.
However, as temperature increases towards $T_c$, the spectral weight of the edge
drops in favor of the broad atomic scattering continuum, characteristic of the ideal (normal) Fermi gas.
This fading of the pair-breaking edge consequently reduces the sensitivity $\sigma$.

In theory, the vertical tangent of $\omega\mapsto\chi_{nn}(\omega,\Delta)$ in $\omega_{\rm th}^+$, 
causes a divergence\footnote{To understand this divergence
and also to obtain rigorous numerical results for the blue curve in Fig.~\ref{sensibilite}, we write
$\chi_{nn}(\omega,\Delta)=\Theta(\omega-\omega_{\rm th}(\Delta))f(\omega-\omega_{\rm th}(\Delta),\Delta)$
where the function $\delta\omega\mapsto f(\delta\omega,\Delta)$ is defined on $[0,+\infty[$. Taking the derivative
with respect to $\Delta$ (and using $f(0,\Delta)=0$), we have $\partial\chi_{nn}/\partial\Delta=\partial f/\partial\Delta-(\dd\omega_{\rm th}/\dd\Delta)\times(\partial f/\partial\delta\omega)$.
The vertical tangent of $\chi_{nn}$ in $\omega_{\rm th}^+$ implies a divergence of $\partial f/\partial\delta\omega$ in $\delta\omega=0^+$, and therefore a divergence of the sensitivity.} 
of the sensitivity at the edge (see the blue curve in Fig.~\ref{sensibilite}), which should favor the detection of Higgs oscillations.
In practice, the finite duration of the Bragg pulse limits the spectral resolution
on $\chi_{nn}$, thereby limiting the maximal accessible value of the sensitivity
to roughly $\sigma \lesssim 2$ (red curve in Fig.~\ref{sensibilite}). This is obtained by broadening $\chi_{nn}$ according to the convolution formula
\be
\chi_{nn}^{\rm br}(\omega_0)=\int_{-\infty}^{+\infty}\dd\omega \chi_{nn}(\omega) f_{\delta\omega}(\omega_0-\omega)
\ee
where $f_{\delta\omega}(\omega)$ is a broadening function parametrized by its energy-width $\delta\omega$. In the comparison performed 
in Fig.~\ref{fig:St_n0p955} we have taken 
$\delta\omega/\epsilon_F=\tau_{\rm F}/t_{\rm B}\simeq0.55$, using the Fermi time $\tau_{\rm F}\simeq27.4402 \mu s$ associated to the density in the trap center at unitarity,
and a Gaussian broadening profile:
\be
 f_{\delta\omega}(\omega)=\frac{1}{\sqrt{2\pi}\delta\omega}\eee^{-\omega^2/2(\delta\omega)^2}.
\ee
However the choice of a Gaussian broadening profile is not crucial for this discussion. Compared to our previous
experimental schemes \cite{Vale2017}, the spectral width is relatively larger here, due to the necessity
of maintaining a Bragg duration much smaller than $2\pi/\omega_{\rm H}$. As visible in Fig.~\ref{sensibilite},
the selected excitation frequency $\omega=\hbar^2q^2/4m$ is fairly optimal for the broadened sensitivity whereas $\omega=\omega_{\rm th}$ is preferable in the absence of broadening.

\begin{figure}[htb]
\begin{center}
\includegraphics[width=0.7\textwidth]{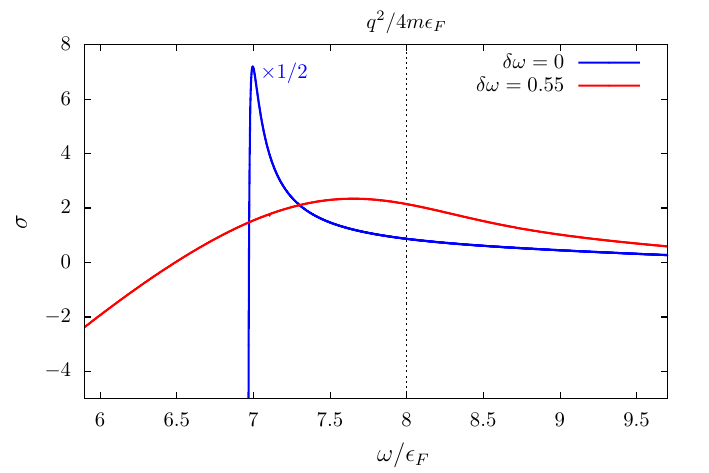}
\caption{\label{sensibilite} The bare (blue curve) and broadened (red curve) sensitivities as a function
of the excitation frequency for $q=4k_F$, $T=0$, and $1/k_F a=0$.}
\end{center}
\end{figure}

Using the sensitivity \eqref{sigma}, we directly predict the scaled Bragg response $S(t)/S_\infty$ from
 \begin{equation}
\frac{\braket{\bar \chi_{nn}''(t)}}{\bar \chi_{nn}''(+\infty)}=1+ \sigma_{f} \frac{\Delta_\infty}{\Delta_f} \left(\frac{\braket{\bar\Delta(t)}}{\Delta_\infty}-1\right),
\label{eq:Sinfty_ave}
\end{equation}
where we have assumed that the width
of the Bragg laser $\lambda$ is small enough such that  the dependence of the sensitivity $\sigma_f\equiv\sigma(\Delta_f)$ on $\rr$ can be neglected. 
Note that the linearization of $\chi_{nn}$ performed in Eq.~\eqref{eq:Sinfty_ave} is valid only in the small amplitude approximation, 
hence at low $T/T_c$.

In Fig.~\ref{fig:St_n0p955}, we compare the theoretical curve Eq.~\eqref{eq:Sinfty_ave} to the experimental measurements of $S(t)/S_\infty$. We consider the lowest temperature case, $T=0.10 T_F$, for which $\delta\Delta$ is small enough with respect to $\Delta_i$ to justify the use of the small-amplitude regime (we have $|\Delta_\infty - \Delta_i| \approx 0.1 \Delta_i$). At the latest times considered in the experiments, we find a small deviation of our theoretical prediction from the homogeneous $1/t^{1/2}$ power-law damping, with $\gamma$ increasing to $0.645\pm 0.005$. This deviation could be seen if we exclude the first peak of the oscillation from the theoretical fitting window, consistently with our analysis in Sec.~\ref{sec:LDA} (compare with Fig.~\ref{fig:peaks} and the inset of Fig.~\ref{fig:St_n0p955}). Instead, using the experimental fitting window, which also includes the first peak, this deviation is nearly compensated, as discussed in the main text. 

\begin{figure}
\centering
\includegraphics[width=0.5\columnwidth]{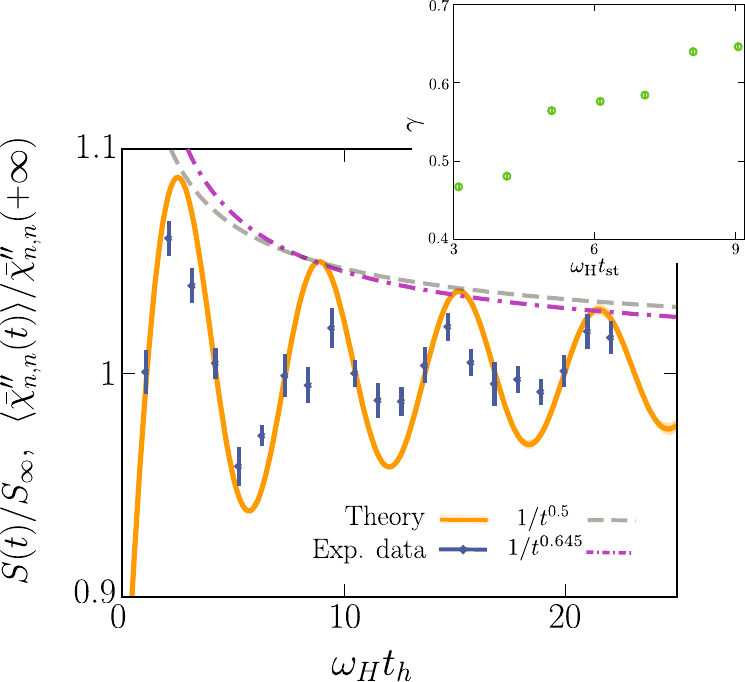}
\caption{Comparison between the experimental Bragg signal $S(t)/S_\infty$ and its theoretical prediction Eq.~\eqref{eq:Sinfty_ave}. The theoretical curve has been obtained at fixed $T=0.66 T_{c, i}$ by including a linear ramp in $50\mu s$ with $t_h\ = t-t_q/2$ corresponding to the theoretical delay found in Sec.~\ref{sec:ramp-rate}, the inhomogeneous broadening, the Bragg time averaging, and the Fourier broadening. The experimental points correspond to $T=0.10 T_F$  and the mean density is $\bar{n} =(0.955\pm 0.018) n_0$ (see main text). We show the deviation from the $1/t^{0.5}$ power-law damping (grey dashed line). The power-law $1/t^{0.645\pm 0.005}$ (dashed-dotted purple curve) has been obtained by fitting the theoretical curve in the time range $9\leq\omega_Ht_h\leq23$ (from the second peak to the last experimental point). Inset: Damping coefficient as a function of the starting time $t_\mathrm{st}$ of the fitting window used for the theoretical curve. }
\label{fig:St_n0p955}
\end{figure} }

\bibliographystyle{apsrev4-1}
\bibliography{supp_biblio}